\documentclass[prx,superscriptaddress,amsfonts,amssymb,amsmath,floats,twocolumn,aps,footinbib,shownopacs]{revtex4}
\usepackage[pdftex]{graphicx}
\usepackage{subfigure}
\usepackage{dsfont}
\usepackage{lipsum}
\usepackage{dcolumn}
\usepackage{bm}
\usepackage{color}
\usepackage{physics}
\usepackage{multirow}
\usepackage[pdftex,colorlinks=true, linkcolor=blue,citecolor=blue,filecolor=blue]{hyperref}
\usepackage[bb=boondox]{mathalfa}

\usepackage{changes}

\begin{document}
\title{Nonthermal order by disorder}

\author{Francesco Grandi} 
\affiliation{Institute for Theory of Statistical Physics, RWTH Aachen University, 52056 Aachen, Germany}
\affiliation{Institut f\"ur Theoretische Physik und Astrophysik and W\"urzburg-Dresden Cluster of Excellence ct.qmat, Universit\"at W\"urzburg, 97074 W\"urzburg, Germany} 
\author{Antonio Picano} 
\affiliation{JEIP, UAR 3573 CNRS, Coll\`{e}ge de France, PSL Research University, 11 Place Marcelin Berthelot, 75321 Paris, France}
\author{Ronny Thomale}
\affiliation{Institut f\"ur Theoretische Physik und Astrophysik and W\"urzburg-Dresden Cluster of Excellence ct.qmat, Universit\"at W\"urzburg, 97074 W\"urzburg, Germany} 
\author{Dante M. Kennes} 
\affiliation{Institute for Theory of Statistical Physics, RWTH Aachen University, 52056 Aachen, Germany}
\affiliation{JARA-Fundamentals of Future Information Technology, 52056 Aachen, Germany}
\affiliation{Max Planck Institute for the Structure and Dynamics of Matter, Center for Free-Electron Laser Science (CFEL), Luruper Chaussee 149, 22761 Hamburg, Germany}
\author{Martin Eckstein}
\affiliation{Institute of Theoretical Physics, University of Hamburg, 20355 Hamburg, Germany}
\affiliation{The Hamburg Centre for Ultrafast Imaging, Hamburg, Germany}

\begin{abstract}
The quench dynamics of systems exhibiting cooperative or almost competitive orders in equilibrium are explored using Ginzburg-Landau theory plus fluctuations. We show that when the renormalization of the free energy by fluctuations is taken into account, anisotropic stiffnesses and relaxation rates of the order parameters can lead to a  stabilization of ordered states at transient free energy minima which are distinct from any (global or local) minima of the equilibrium free energy. This theory demonstrates that nonequilibrium fluctuations play a pivotal role in forming nonthermal orders. As nonthermal order and nonthermal fluctuations mutually stabilize each other over some time, this mechanism could be seen as a nonequilibrium variant of the order-by-disorder phenomenon. We discuss the potential relevance of these findings for systems with intertwined orders, such as superconductivity and density wave orders, relevant for high-temperature superconductors and the kagome metals, as well as for systems that show orbital ordering.
\end{abstract}

\maketitle

\section{Introduction}
The long-term goal of ultrafast science is to achieve nonequilibrium and ultrafast control over materials' properties \cite{Giannetti2016_AdvPhys,Basov2017_NatMat,delaTorre2021_RMP,Murakami2023_arXiv}. Key direct control pathways include Floquet engineering, where tailored periodic pump pulses are employed to modify the electronic band structure \cite{McIver2020_NatPhys,Oka2009_PRB}, and nonlinear phononics, which enables selective excitation of specific phonon modes to manipulate both electronic properties and the crystal structure \cite{Foerst2011_PRB,Mankowsky2014_Nat,Nova2017_NatPhys,Mankowsky2017_PRL,vonHoegen2018_Nat}. While these methods are effective during the active application of perturbations, many experiments reveal phenomena that persist long after the excitation is removed. Among these, the most striking is the emergence of metastable states which cannot be reached along quasi-static thermodynamic pathways -- commonly referred to as hidden states -- after the action of an incoherent excitation. Examples include photo-induced magnetic and orbital orders \cite{Ichikawa2011_NatMat}, the metal-insulator transition in the quasi-two-dimensional compound $1T$-TaS$_2$ \cite{Stojchevska2014_Science}, and metastable light-induced superconductivity (SC) \cite{Budden2021_NatPhys}.

The mechanisms underlying the transient stabilization of nonthermal orders remain poorly understood. One possible microscopic origin involve electronic relaxation bottlenecks, which may lead to long-lived photo-doped phases \cite{Murakami2023_arXiv}. At a phenomenological level, valuable insights into control pathways have been gained by examining the dynamics within complex  free energy landscapes. In such cases, pump pulses can apply effective forces to the order parameter, enabling the system to be steered or coherently controlled along specific routes in the free energy landscape \cite{Forst2011_NatPhys,Gassner2024_NatComm,Maklar2023_SciAdv}, in particular in systems with multiple order parameters that respond differently to external perturbations \cite{Tokura2006_JPSJ,Zhang2016_NatMat,Teitelbaum2019_PRL}. Moreover, it has been noted that the transition from a global minimum to a hidden phase is a dynamical phenomenon which depends on different relaxation times of intertwined orders rather than only on the equilibrium free energy alone \cite{Sun2020_PRX}. The above discussion of excitation pathways relies on the assumption of metastable minima existing in the equilibrium free energy in some form. In this work, we will demonstrate that in a genuine nonequilibrium situation a transient stabilization of nonthermal orders can arise even if the equilibrium free energy has no metastable states to begin with.

\begin{figure*}
    \centerline{\includegraphics[width=0.98\textwidth]{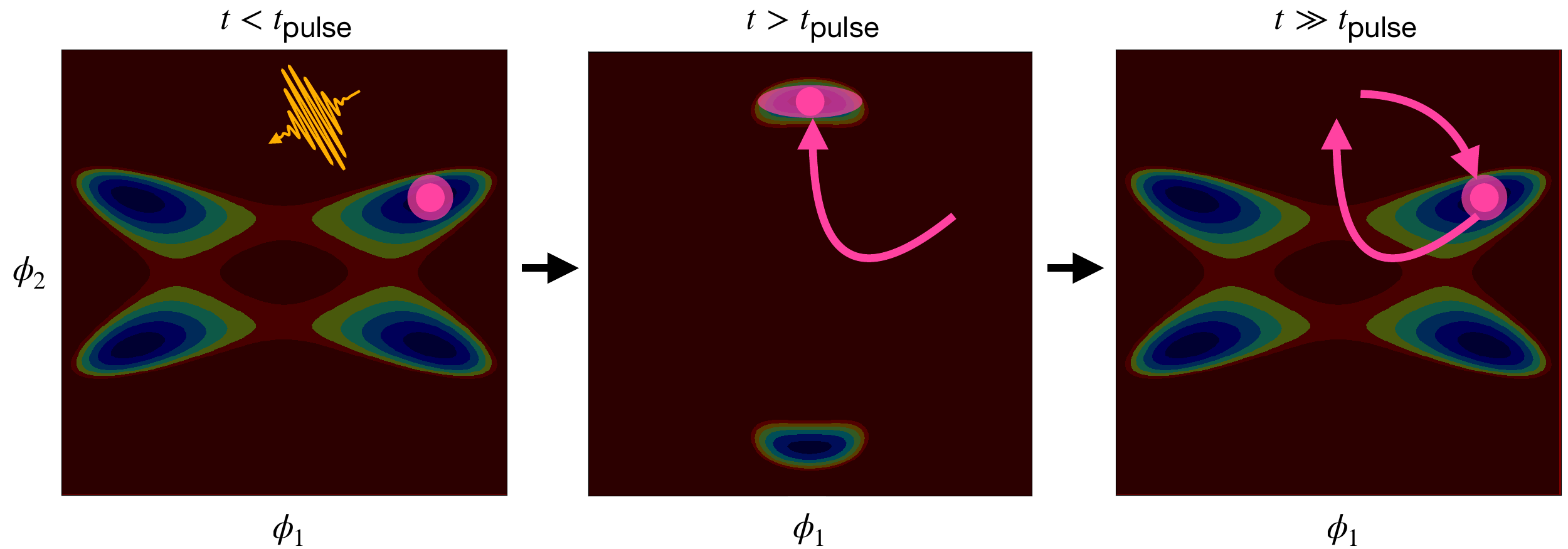}}
	\caption{ \textbf{Nonthermal order by disorder (NOBD) -} Instead of steering an order parameter $\phi=(\phi_1,\phi_2)$ through a free energy landscape with existing metastable minima, a new minimum (nonthermal order) is transiently generated by activating a nonthermal and possibly anisotropic distribution of spatial fluctuations of $\phi$ (illustrated by the ellipse-shade surrounding the circle). Through anharmonic interactions of the order parameter and its fluctuations they can mutually stabilize each other in the nonthermal state for some time.} \label{fig:scheme}
\end{figure*}

In general, the temperature-dependent free energy $F(T, \bar{\phi})$ of a coarse-grained order parameter $\bar{\phi}$ can be conceptualized as arising from the elimination of other degrees of freedom, assuming these remain approximately thermalized. This assumption is often valid for certain ``fast degrees of freedom" (e.g., electrons) and can be refined through multi-temperature models to better capture the dynamics of the order parameter \cite{Perfetti2007_PRL, Johnson2017_StruDyn, Maklar2021_NatComm, Dolgirev2020_PRB2}. However, the order parameter $\phi(\mathbf{r}) = \bar{\phi} + \delta\phi(\mathbf{r})$ exhibits spatial fluctuations, and the thermalization of these fluctuations $\delta\phi(\mathbf{r})$ becomes increasingly slow at longer length scales \cite{Zong2019_NatPhys, Kogar2020_NatPhys, Zong2021_PRL}. The effect of these  nonthermal fluctuations on the dynamics can be significant: In a rotationally invariant $\phi^4$ theory, the interaction between the fluctuations and the order parameter leads to an effective softening of the free energy potential $F(T, \bar{\phi})$  and a considerable slow-down of the dynamics \cite{Dolgirev2020_PRB}. Here, we argue that nonthermal fluctuations can also renormalize the free energy landscape in such a way that entirely new order parameter configurations are stabilized. The resulting nonthermal order, in turn, alters the distribution of the fluctuations, leading to a mutual stabilization of nonthermal order and nonthermal fluctuations over some period of time (Fig.~\ref{fig:scheme}).

The contribution to the dynamics played by the nonthermal fluctuations is expected to be particularly relevant in systems that, already in equilibrium, have a free energy landscape strongly renormalized by the order parameter fluctuations. This is the case of the systems that go through an order-by-disorder (OBD) transition \cite{Villain1980_JPF,Shender1982_JETP,Henley1989_PRL,Moessner2001_CJP,Nussinov2015_RMP,Lenz2024_CommPhys}. Moreover, fluctuations are expected to play a relevant role even for vestigial states of matter, as might be the case for the nematicity observed in several classes of compounds \cite{Fernandes2012_PRB,Svistunov2015_SSM,Fernandes2019_ARCMP,Grandi2024_PRB}. In particular, the OBD mechanism leads to a renormalization of the free energy due to quantum \cite{Rastelli1987_JPC,Brink2004_NJP} (at zero temperature) or thermal \cite{Sachdev1992_PRB,Biskup2005_CommMatPhys} (at finite temperature) fluctuations, effectively creating new minima in the energy landscape. Out of equilibrium, the relevance of the fluctuations in the dynamics of the order parameter might extend beyond these classes of systems. Indeed, nonequilibrium fluctuations lead to inhomogeneous disordering in several correlated materials \cite{Wall2018_Science,PerezSalinas2022_NatComm,Johnson2022_PRL}, a signature of which might be given by a transient multimodal distribution of the order parameters \cite{Picano2023_PRB}.

In the following, we focus on systems that show cooperative orders, i.e., in which two order parameters $\phi_1$ and $\phi_2$ are simultaneously different from zero in equilibrium, or on systems that show an almost competitive order (one of the order parameters is significantly larger than the other, $\phi_1 > \phi_2 \sim 0$). Both conditions are relevant for different classes of materials, ranging from the cuprates \cite{Fradkin2015_RMP}, the iron-based superconductors \cite{Pratt2009_PRL,Fernandes2010_PRB}, the vanadium-based kagome metals (where unconventional SC coexisting with a charge-ordered state has been observed) \cite{Neupert2022_NatPhys,Wilson2024_NatRevMat}, and orbitally ordered compounds such as KCuF$_3$ \cite{Pavarini2008_PRL,Li2018_NatComm}.

Since the renormalization of the free energy by fluctuations is important also in equilibrium, it is interesting to contrast this to the nonthermal order by disorder (NOBD) mechanism that will be discussed in this manuscript: The free energy functional $\mathcal{F}$ of the system has a local contribution $\mathcal{V}$ which includes the interaction among the fields, and a nonlocal stiffness contribution $\mathcal{K}$ that takes into account the energy cost of creating spatially anisotropic fields, $\mathcal{F} = \mathcal{V} + \mathcal{K}$. In equilibrium, an effective free energy potential $F(\bar \phi_1,\bar \phi_2,T)$ can be computed by taking into account the {\em thermal} fluctuations around the configuration $(\bar \phi_1, \bar \phi_2)$. Thus, a state with a strong entropic contribution coming from the spatial fluctuations might become a minimum of the free energy of the problem even if, without the fluctuations, would not be stable. This is the so-called entropic stabilization, or OBD, mechanism. Out of equilibrium, and particularly on short time scales, the excited fluctuations are nonthermal, and their magnitude and distribution depend on the specific excitation path the system went through. Nevertheless, one can still get the effective free energy which is seen by the time dependent order parameters $(\phi_1 (t), \phi_2 (t))$ as renormalized by the nonthermal fluctuations at a given time $t$ of the dynamics. This renormalized free energy can have new minima in particular when the two order parameters have some form of anisotropy which might be dynamical (different relaxation rates), spatial (the distribution of the fluctuations is spatially anisotropic) or intrinsic (the stiffnesses $\mathcal{K}$ of the two order parameters are different). The effect is present only if the renormalization of the energy landscape due to the fluctuations of the order parameter is taken into account. The phenomenon can therefore be regarded as a nonequilibrium version of the OBD mechanism (NOBD), which does not require OBD to be present in the equilibrium state \cite{Wan2017_PRL} and is not limited to a specific condition \cite{Zhou2024_arXiv}. In other words, it is a general mechanism that might be relevant for various experimental setups.

\section*{Results}
\subsection*{Equilibrium free energy}
We consider a system with two order parameters $\phi_1$ and $\phi_2$. For all situations considered below we assume that the system has a $\mathbb{Z}_2 \times \mathbb{Z}_2$ symmetry, defined by a mirror symmetry $\phi_\alpha\to-\phi_\alpha$ in both components. A generic Landau potential reads 
\begin{align} \label{eq:pot_z2xz2_c4}
	\mathcal{V} (\phi_1, \phi_2) & = \frac{1}{2} \sum_{\alpha} r_\alpha (T) \phi_\alpha^2 + \frac{u_1}{4} (2 \phi_1 \phi_2 )^2 \nonumber \\
    & + \frac{u_2}{4} (\phi_1^2 - \phi_2^2)^2,
\end{align}
where  $r_\alpha (T) = r_{\alpha,0} ( \frac{T}{T_{\text{c}}} - 1)$ shows the  temperature dependence of the quadratic coefficients close to the critical temperature $T_{\text{c}}$, and $r_{\alpha,0} > 0$. To reduce the number of parameters, we have written the potential such that the quartic part has a higher symmetry (C$_4$), which is sufficient to discuss the qualitative phenomena below. The contribution of $\mathcal{V}$ that is proportional to $u_1$ tends to stabilize a state with $\phi_1 = 0$ or $\phi_2 = 0$, while a large $u_2$ would favor $\phi_1 = \phi_2$ if $r_1=r_2$. Below, we focus on the latter case, $u_2 > u_1$. With this choice, we can still discuss two qualitatively different situations, which will simply be referred to as cooperative and competitive order below: (i) The case of cooperative order is obtained if both $r_{1,0}/r_{2,0}$ and $r_{2,0}/r_{1,0}$ are less than  $u_2 / (2 u_1 - u_2)$. For $T < T_\text{c}$, one then finds four degenerate minima of $\mathcal{V}$ in which both order parameters are simultaneously nonzero,
\begin{align}
    \phi_\alpha (T) = \pm \sqrt{\frac{2 u_1 r_{\bar{\alpha}} (T) - u_2 \big( r_1 (T) + r_2 (T) \big)}{4u_1 (u_2 - u_1)}} \nonumber
\end{align}
(here $\bar\alpha=2$ for $\alpha=1$ and vice versa). (ii) The case of competitive order is obtained for $r_{\alpha,0}/r_{\bar{\alpha},0} \ge u_2 / (2 u_1 - u_2)$.  For $T < T_\text{c}$, there are then only two degenerate minima at $\phi_\alpha (T) = \pm \sqrt{- r_\alpha (T) / u_2}$, while the subdominant $\phi_{\bar{\alpha}}$ order parameter vanishes. We emphasize that in neither of the cases the potential has metastable minima.

As the aim of this work is to understand the effect of nonthermal spacial fluctuations of the order parameters, we allow for  spatially fluctuating order parameters $\phi_1(\bm r)$ and $\phi_2(\bm r)$, and extend the free energy to
\begin{align} \label{eq:free_en}
    \mathcal{F} [\phi_1, \phi_2] = \int d^d r \ \big[ \mathcal{V} ( \phi_1, \phi_2 ) + \mathcal{K} (\phi_1, \phi_2) \big] ,
\end{align}
where $d$ is the dimensionality, and $\mathcal{K} (\phi_1, \phi_2)$ represents the stiffness. The latter is a general symmetry allowed quadratic form in the spatial gradients $\partial_i \phi_\alpha$
\begin{align} \label{eq:stiff_gen}
    \mathcal{K} (\phi_1, \phi_2) & = \frac{1}{2} \sum_\alpha \sum_i \tilde{K}_{\alpha, i} (\partial_i \phi_\alpha)^2
\end{align}
(here $i=1,\ldots,d$ labels the spatial dimensions). The precise form of the stiffness term is specified in the examples below. The stiffness controls both the equilibrium distribution and the dynamics of order parameter fluctuations, which will then renormalize the free energy seen by the coarse-grained order parameters.

\subsection*{Dynamics}
We let the order parameter fields evolve in time according to relaxational time-dependent Ginzburg-Landau (model A) dynamics \cite{Hohenberg1977_RMP,Tauber2014_book}. This leads to the Langevin equation
\begin{align}
    & \partial_t \phi_\alpha (\mathbf{r},t) = - \Gamma_\alpha \frac{\delta \mathcal{F}}{\delta \phi_\alpha (\mathbf{r},t)} + \eta_\alpha (\mathbf{r},t), \nonumber
\end{align}
where $\Gamma_\alpha > 0$ is the relaxation rate for the order parameter $\phi_\alpha (\mathbf{r},t)$, and $\eta_\alpha (\mathbf{r},t)$ is a Gaussian white noise with zero mean $\langle \eta_\alpha (\mathbf{r},t) \rangle  = 0$, which fulfills the fluctuation-dissipation theorem
\begin{align} \label{eq:noise_distrib}
    & \langle \eta_\alpha (\mathbf{r},t) \eta_\beta (\mathbf{r}',t') \rangle = 2 \Gamma_{\alpha} T \delta (\mathbf{r} - \mathbf{r}') \delta (t - t') \delta_{\alpha \beta} .
\end{align}
As the system has only an overall $\mathbb{Z}_2 \times \mathbb{Z}_2$ symmetry, we will later allow for a difference in the relaxation rates $\Gamma_\alpha$ for the two order parameters.

The incoherent laser-induced excitation is modeled by a transient increase of the temperature of the fast degrees of freedom, such as the electrons, which are assumed to thermalize instantaneously on the time scale of the order parameter dynamics. Specifically, we will assume a simple time-dependent temperature profile,
\begin{align} \label{eq:temp_quench}
    T (t) = T_\text{i} + (T_\text{q} - T_\text{i}) \theta (t - t_\text{q}) e^{-(t-t_\text{q})/\tau} ,
\end{align}
given by a step-like increase of the temperature (quench) from the initial temperature $T_\text{i}$ to an excited temperature $T_\text{q}$ at time $t_\text{q}$, and the recovery of the initial temperature $T_\text{i}$ with a relaxation time $\tau$. Besides this incoherent excitation, we will also consider the possibility of a more coherent control protocol, which will be described later.

To solve the equations, we make a Gaussian approximation for the probability distribution for the fields $\phi_\alpha (\mathbf{r},t)$. The assumptions of this theoretical approach has been already applied to the description of some experimental properties of tritelluride materials \cite{Zong2021_PRL,Zong2019_NatPhys}. Within the Gaussian approximation, one can derive a closed set of equations for the average order parameters $\bar{\phi}_\alpha (t) = \langle \phi_\alpha (\mathbf{r}, t) \rangle$ and for the correlation functions $D_{\alpha \beta} (\mathbf{k}, t) = \langle \phi_\alpha (\mathbf{k}, t) \phi_\beta (- \mathbf{k}, t) \rangle$ defined in reciprocal space (see the \textit{Methods} for the explicit expression of the equations). In particular, the average order parameters satisfy the equation
\begin{align} \label{eq:av_op_main_text}
    \dot{\bar{\phi}}_\alpha (t) = - \Gamma_\alpha \sum_{\beta = 1}^2 
   ( \bar r_{\alpha \beta} +    r^\text{fl}_{\alpha \beta})
    \bar{\phi}_\beta (t) ,
\end{align}
where  the effective force constant
\begin{align}
    & \bar r_{\alpha \beta} = \big[ r_\alpha (T) + u_2 \bar{\phi}_\alpha^2 + (2 u_1 - u_2) \bar{\phi}_{\bar{\alpha}}^2  \big] \delta_{\alpha \beta} \nonumber 
\end{align}
includes the anharmonic self-interaction of $\bar \phi$ through the fourth order terms in $\mathcal{V}$, and in addition there are corrections 
\begin{align}
    r_{\alpha \beta}^\text{fl} = & \big[ (2 u_1 - u_2) (n_{\bar{\alpha} \bar{\alpha}} - 2 n_{\alpha \beta}) + 3 u_2 n_{\alpha \alpha} \big] \delta_{\alpha \beta} \nonumber \\
    & + 2 (2 u_1 - u_2) n_{\alpha \beta} . 
    \label{eq:Rfluct}
\end{align}
which arise from the interaction of $\bar \phi$ with fluctuations; the latter enter through the integrated contribution
\begin{align} \label{eq:fluct}
    n_{\alpha \beta} (t) = \int^{\Lambda_\text{c}} \frac{d \mathbf{k}}{(2 \pi)^d} \ D_{\alpha \beta} (\mathbf{k}, t)
\end{align}
in each channel ($\Lambda_{\text{c}}$ is an ultraviolet cutoff).  During the dynamics, the fluctuations $n_{\alpha \beta} (t)$ evolve in a nontrivial way and thus renormalize the dynamics of the average order parameter.

\subsection*{Effective potential}
In order to quantify and analyze the effect of the fluctuations on the dynamics, one can introduce an effective potential $\bar F$ for the average order parameter $\bar{\phi}_\alpha$ in the presence of fluctuations; $\bar F$ will be defined such that its derivative reproduces the equations of motion for $\bar{\phi}_\alpha$:
\begin{align}
    \dot{\bar{\phi}}_\alpha  (t) = - \Gamma_\alpha \frac{\partial \bar{F} (\bar{\phi}_1, \bar{\phi}_2)}{\partial \bar{\phi}_\alpha}. \nonumber
\end{align}
By comparison with Eq.~\eqref{eq:av_op_main_text}, one finds
\begin{align} \label{eq:free_en_fluct}
    \bar{F} (\bar{\phi}_1, \bar{\phi}_2) = \bar{F}^0 (\bar{\phi}_1, \bar{\phi}_2) + \bar{F}^\text{fl} (\bar{\phi}_1, \bar{\phi}_2),
\end{align}
where  $\bar{F}^0 (\bar{\phi}_1, \bar{\phi}_2)= \mathcal{V}(\bar{\phi}_1, \bar{\phi}_2)$ is the unrenormalized potential Eq.~\eqref{eq:pot_z2xz2_c4}, and the fluctuations correction $\bar{F}^\text{fl} (\bar{\phi}_1, \bar{\phi}_2)$ can be written in the form 
\begin{align}
    \bar{F}^\text{fl} (\bar{\phi}_1, \bar{\phi}_2) = \frac{1}{2} \sum_{\alpha \beta} \bar{\phi}_\alpha r_{\alpha \beta}^\text{fl} \bar{\phi}_\beta ,
\end{align}
with the fluctuation force constants \eqref{eq:Rfluct}. The dynamics of $\bar{F} (\bar{\phi}_1, \bar{\phi}_2)$ provides a convenient tool to understand and quantify the dynamics of the averages $\bar{\phi}_1$ and $\bar{\phi}_2$.

For later convenience, we introduce the polar coordinates $R$ (amplitude) and $\varphi$ (phase) to parametrize the dynamics of the order parameters $\bar{\phi}_1$ and $\bar{\phi}_2$:
\begin{align}
    & \bar{\phi}_1 = R \cos (\varphi) , \nonumber \\
    & \bar{\phi}_2 = R \sin (\varphi) . \nonumber
\end{align}
In particular, a value of $\varphi = 0$ ($\varphi = \pi/2$) corresponds to an order parameter that has only the component $\bar{\phi}_1$ ($\bar{\phi}_2$). If, instead, $\varphi = \pi/4$, the two components of the order parameter are equal among them, $\bar{\phi}_1 = \bar{\phi}_2$.

\begin{figure}
    \centerline{\includegraphics[width=0.5\textwidth]{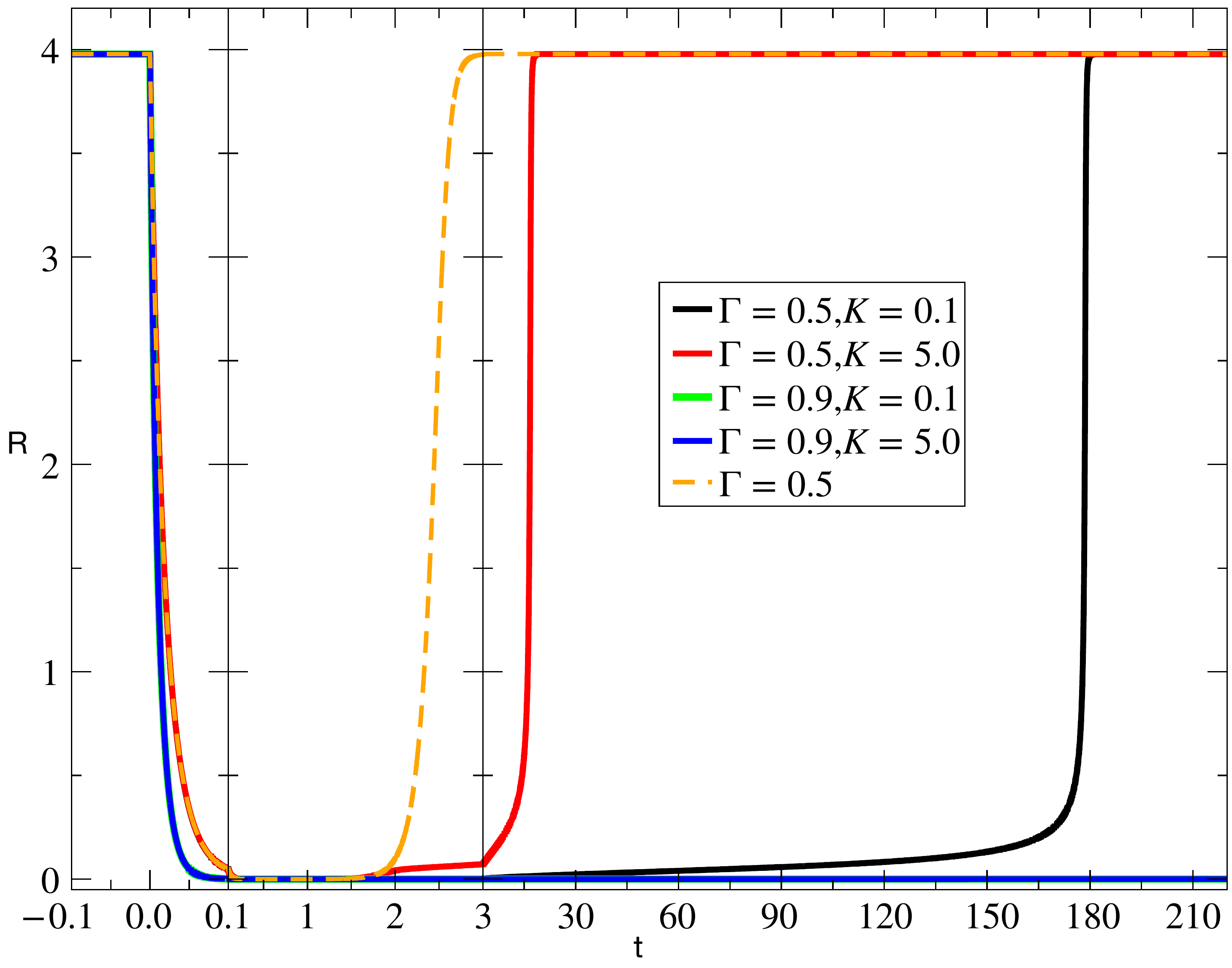}}
	\caption{ { \textbf{Amplitude dynamics after a temperature quench (isotropic fluctuations) -} Amplitude $R$ of the order parameter after a temperature quench. We contrast the dynamics for several combinations of $\Gamma$ and $K$ with the dynamics in the absence of the fluctuations (orange dashed line with $\Gamma = 0.5$). In the absence of fluctuations, the recovery of the original amplitude $R \sim 4$ occurs on a faster time-scale than when the fluctuations are included. At fixed $\Gamma$, the recovery takes place on a shorter time-scale at larger $K$. For higher values of $\Gamma$, the recovery happens at longer times (for $\Gamma = 0.9$, both the green and the blue curves are very close to $R = 0$ even for the longest time considered in our simulation). The horizontal time-axis emphasizes the short (left), intermediate (middle) and long (right) time dynamics. The parameters of the temperature quench Eq.~\eqref{eq:temp_quench} are $T_\text{i} = 0.025$, $T_\text{q} = 4$, $t_\text{q} = 0$, $\tau = 0.3$; the parameters for the potential are $T_\text{c} = 0.5$, $r_0 = 15$, $u_1 = 0.9$, $u_2 = 1.0$, $d = 3$ and $\Lambda_\text{c} = 2 \pi$.} } \label{fig:rad_vs_time_sym}
\end{figure}

\subsection*{Isotropic fluctuations}
As a first illustration of the effect of fluctuations on the dynamics, we consider a case of higher symmetry. For this purpose, we take the potential Eq.~\eqref{eq:pot_z2xz2_c4} to be C$_4$-symmetric, i.e., $r_{1,0} = r_{2,0} \equiv r_0$:
\begin{align} \label{eq:pot_c4}
	\mathcal{V} (\phi_1, \phi_2) & = \frac{r (T)}{2} \sum_{\alpha} \phi_\alpha^2 + u_1 (\phi_1 \phi_2 )^2 \nonumber \\
    & + \frac{u_2}{4} (\phi_1^2 - \phi_2^2)^2 .
\end{align}
Below the critical temperature, Eq.~\eqref{eq:pot_c4} has four degenerate minima, with $R= \sqrt{- r (T)/u_1}$ and $\varphi\in\{\pm\pi/4,\pm3\pi/4\}$ in the polar representation. We combine this potential with a stiffness contribution that is isotropic both in real space and in the two order parameters,
\begin{align} 
    \mathcal{K} (\phi_1, \phi_2) = \frac{r_0}{2} K \sum_{\alpha} \big( \nabla \phi_\alpha \big)^2, \nonumber
\end{align}
and consider the relaxation rates to be equal for $\phi_1$ and $\phi_2$, $\Gamma_1 = \Gamma_2 = \Gamma$. If the dynamics is started in one of the four degenerate equilibrium states, the symmetry of the problem implies that the phase of the order parameter remains locked to the initial value. For this reason, we will only discuss the dynamics of the absolute value $R$ in this section.

\begin{figure}
    \centerline{\includegraphics[width=0.5\textwidth]{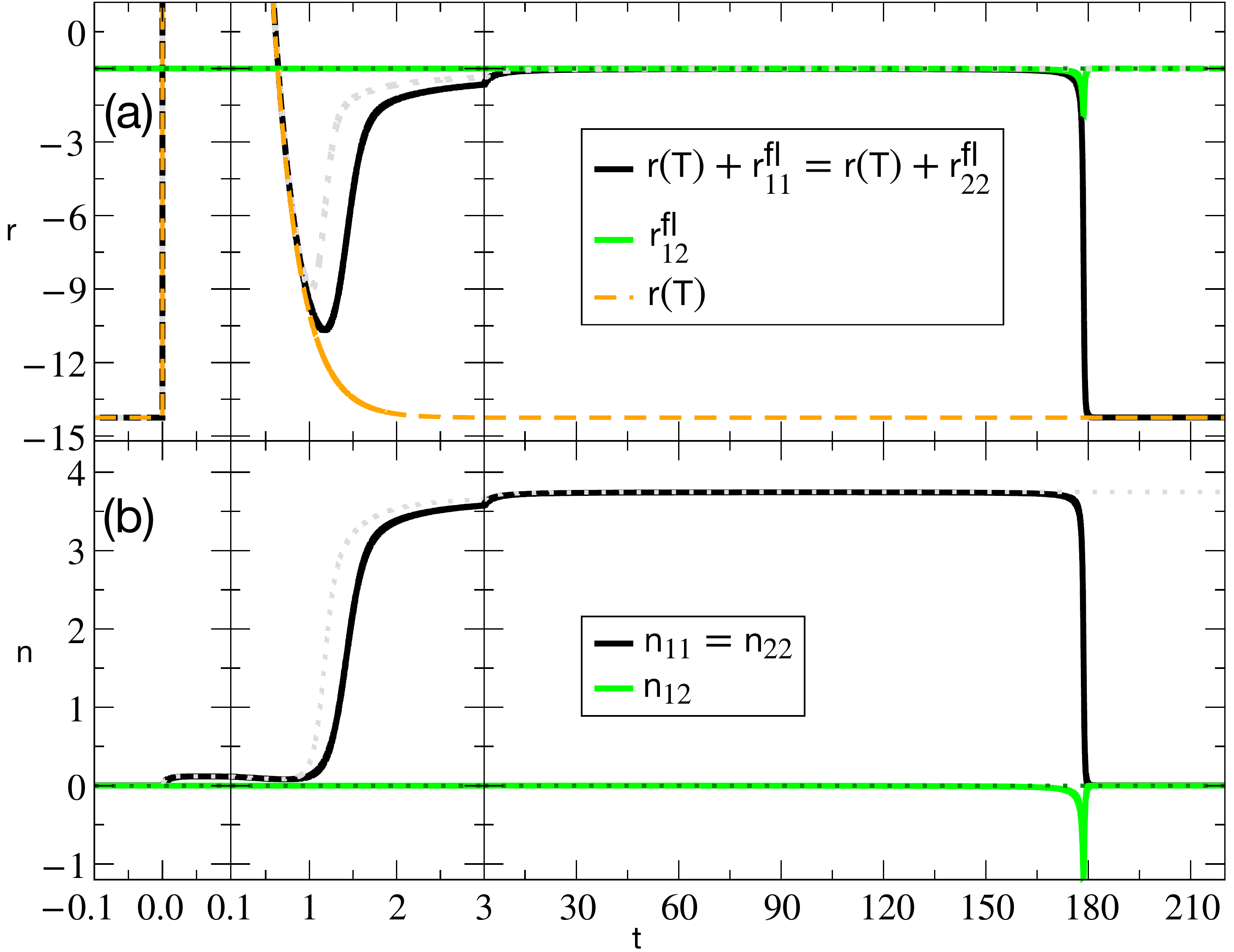}}
	\caption{ { \textbf{Fluctuations dynamics after a temperature quench (isotropic fluctuations) -} Dynamics of the quadratic coefficients renormalized by the fluctuations (a) and of the excitations per channel (b) during the same time-dependent protocol considered in Fig.~\ref{fig:rad_vs_time_sym}. All the lines are computed for $K = 0.1$, and the solid (dotted) lines are characterized by $\Gamma = 0.5$ ($\Gamma = 0.9$). The dotted gray (green) lines represent $r(T) + r^\text{fl}_{11} = r(T) + r^\text{fl}_{22}$ ($r^\text{fl}_{12}$) and $n_{11} = n_{22}$ ($n_{12}$) in (a) and in (b), respectively. The dashed orange line in (a) represents the time-dependent quadratic coefficient unrenormalized by the fluctuations. The horizontal time-axis emphasizes the short (left), intermediate (middle) and long (right) time dynamics.}
    } \label{fig:r_n_vs_time_sym}
\end{figure}

After a temperature quench above the critical temperature $T_\text{c}$, the amplitude of the order parameter $R$ rapidly drops to a value very close to zero, see Fig.~\ref{fig:rad_vs_time_sym}. After the temperature and the bare $r$ parameter return to their initial values $T_{\rm i}$ and $r(T_{\rm i})$ (see Fig.~\ref{fig:r_n_vs_time_sym}(a) for a plot of $r(T)$), the state is close to a potential maximum. If the fluctuation contribution to the dynamics is neglected, the order parameter will therefore exponentially increase with a rate $|\Gamma r|$, and quickly recovers the original equilibrium value (see orange dashed line with $\Gamma = 0.5$ in Fig.~\ref{fig:rad_vs_time_sym}). If instead the fluctuations are included (curves with $K = 0.1$ or $K = 5.0$ in Fig.~\ref{fig:rad_vs_time_sym}), the collapse of the order parameter at short times is largely independent of $K$, but the recovery of the original amplitude can be strongly delayed. Neglecting the fluctuations contribution (or the feedback of the fluctuations on the order parameter) corresponds to setting $r^\text{fl}_{\alpha \beta} = 0$ for every $\alpha$ and $\beta$ in Eq.(6), as well as to neglect the dynamics of the correlation functions. Comparing this case to the one where the fluctuations are taken into account is relevant given the popularity of the former approach \cite{Beaud2014_NatMat,Zong2019_PRL,Maklar2021_NatComm}.

\begin{figure*}
    \centerline{\includegraphics[width=0.98\textwidth]{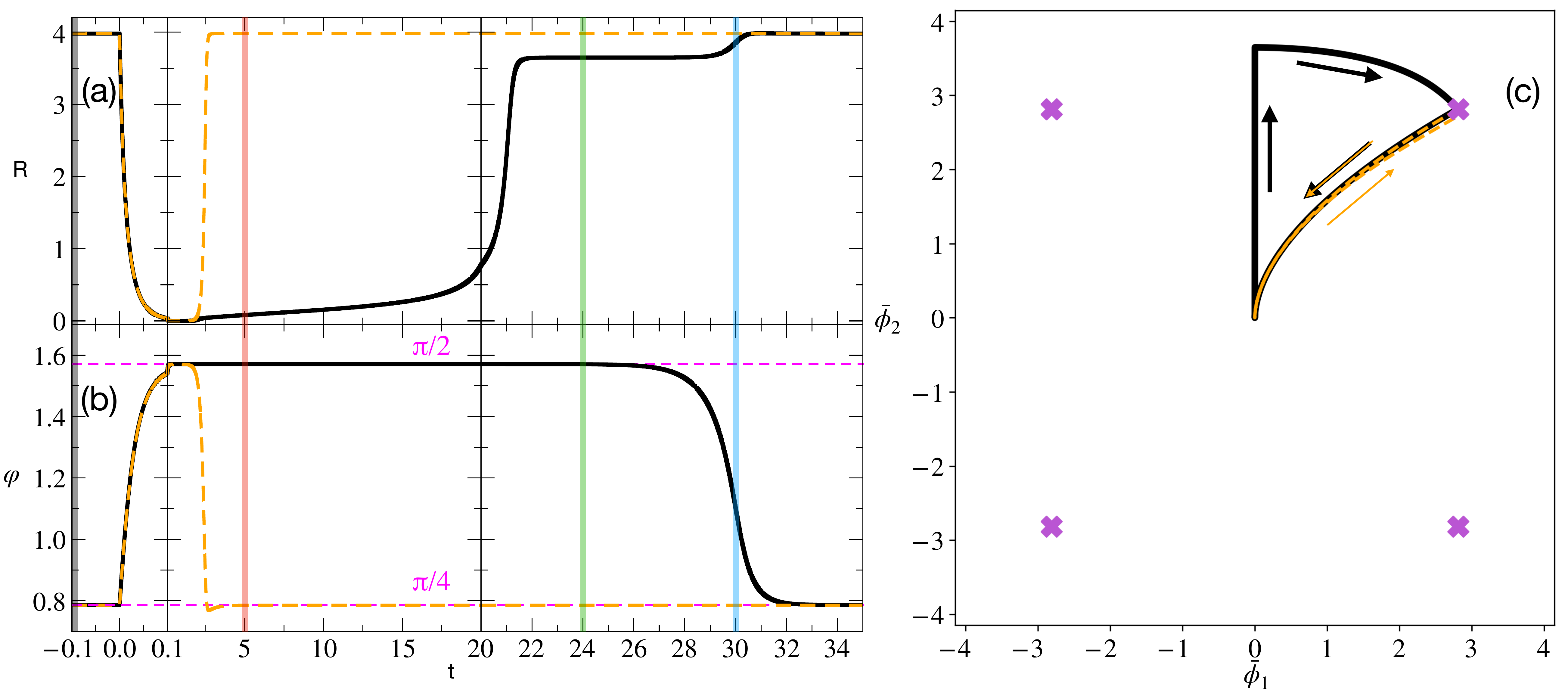}}
	\caption{ { \textbf{Amplitude, phase and trajectory for $\Gamma_1 > \Gamma_2$ and $K_1 < K_2$ after a temperature quench (cooperative order) -} Dynamics of the amplitude $R$ (a) and the phase $\varphi$ (b) of the order parameters after a temperature quench for $\Gamma_1 = 0.9$, $\Gamma_2 = 0.5$, $K_1 = 0.1$ and $K_2 = 5.0$. The dynamics of $R$ shows the emergence of a plateau in the time-window $22 - 30$. The dashed orange lines correspond to the dynamics of $R$ and $\varphi$ without the feedback from the fluctuations. (c) Trajectories drawn by the order parameter in the $(\bar{\phi}_1, \bar{\phi}_2)$ plane during the time-dependent processes shown in panels (a-b). The light-violet crosses represent the position of the equilibrium minima in the renormalized potential. The parameters for the quench and for the equilibrium potential are the same as described in the caption of Fig.~\ref{fig:rad_vs_time_sym}. The shaded vertical lines in panels (a-b) represent the times at which the renormalized free energies are displayed in Fig.~\ref{fig:r_n_vs_time_diff_GAM_AND_K}(c-f).}
    } \label{fig:rad_phi_vs_time_diff_GAM_AND_K_free_en}
\end{figure*}

This behavior can be understood by looking at the dynamics of the fluctuations (see Fig.~\ref{fig:r_n_vs_time_sym}(b)): At the early stage of the dynamics, the fluctuations are small and therefore do not contribute much to the renormalization of the free energy. The fluctuations $n_{\alpha \alpha}$ start to grow rapidly around $t \sim 1$, after $R$ has approached zero, compare Figs.~\ref{fig:r_n_vs_time_sym} and \ref{fig:rad_vs_time_sym}. As a consequence, the effective $r$ parameter $r^{\rm eff} = \bar r+r^{\rm fl}$ almost vanishes in the intermediate time range, see Fig.~\ref{fig:r_n_vs_time_sym}(a). This fluctuation-induced softening of the effective potential explains the slow-down of the dynamics. The effect is more pronounced for larger $\Gamma$ and for smaller $K$, both of which increase the amount of fluctuations: A larger $\Gamma$ implies stronger noise through Eq.~\eqref{eq:noise_distrib}, while a smaller stiffness constant implies both a lower energy cost for creating spatial fluctuations, and a slower decay of the excitations $n_{\alpha \alpha}$.

\begin{figure*}
    \centerline{\includegraphics[width=0.98\textwidth]{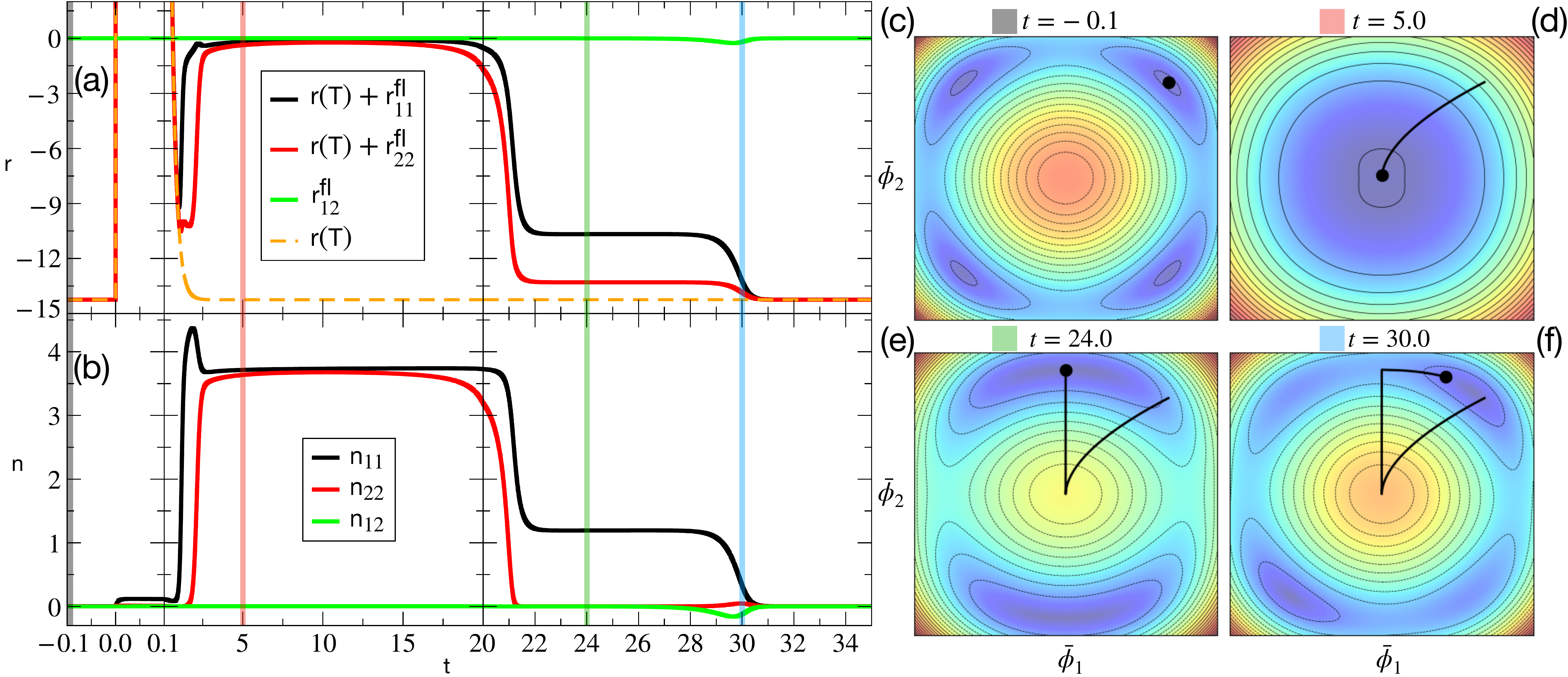}}
	\caption{ { \textbf{Fluctuations and free-energy dynamics for $\Gamma_1 > \Gamma_2$ and $K_1 < K_2$ after a temperature quench (cooperative order) -} Dynamics of the quadratic coefficients renormalized by the fluctuations (a) and of the excitations per channel (b) during the same time-dependent protocol considered in Figs.~\ref{fig:rad_vs_time_sym}-\ref{fig:rad_phi_vs_time_diff_GAM_AND_K_free_en}}. Panels (c-f) show the free energy renormalized by the fluctuations $\bar{F} (\bar{\phi}_1, \bar{\phi}_2 )$, see Eq.~\eqref{eq:free_en_fluct}, at several times before and after the temperature quench. The black dot shows the order parameter at a given time in the $(\bar{\phi}_1, \bar{\phi}_2 )$ plane, and the black line shows the trajectory drawn by the order parameters until the considered time. The parameters are $\Gamma_1 = 0.9$, $\Gamma_2 = 0.5$, $K_1 = 0.1$ and $K_2 = 5.0$. The dashed orange line in (a) represents the time-dependent quadratic coefficient unrenormalized by the fluctuations. The shaded vertical lines in panels (a-b) represent the times at which the renormalized free energies are displayed. For the movie of the process, see the Supplementary Note.}
    \label{fig:r_n_vs_time_diff_GAM_AND_K}
\end{figure*}

\subsection*{$\mathbb{Z}_2 \times \mathbb{Z}_2$ model}
The softening of the potential and the slowdown of the dynamics due to the fluctuations observed in the previous case is qualitatively the same effect as discussed in Ref.~\cite{Dolgirev2020_PRB} for a potential with continuous symmetry, or in Ref.~\cite{Grandi2020_PRR} for a model similar to the one studied here. The main question is now whether nonthermal fluctuations in a system with lower symmetry can also modify the shape of the potential. Here, we show that this can indeed be the case for both the cases of cooperative and competitive order, when the stiffness and/or the relaxation rates are different for the two order parameters, but respect only the $\mathbb{Z}_2 \times \mathbb{Z}_2$ symmetry. Specifically, we assume the potential \eqref{eq:pot_z2xz2_c4} in combination with a stiffness contribution
\begin{align} \label{eq:stiff_z2xz2}
    \mathcal{K} (\phi_1, \phi_2) = \frac{1}{2} \sum_{\alpha} r_{\alpha,0} K_\alpha \big( \nabla \phi_\alpha \big)^2,
\end{align}
with coefficients $\tilde{K}_{i,\alpha} = r_{\alpha,0} K_\alpha$ that can differ for the two orders.

\subsubsection*{Cooperative order}
To discuss the case of cooperative order, we can still let $r_{1,0} = r_{2,0} = r_0$, which makes the potential Eq.~\eqref{eq:pot_z2xz2_c4} C$_4$-symmetric and identical to Eq.~\eqref{eq:pot_c4}. To make the two orders inequivalent, we take a condition where $\bar{\phi}_1$ has a smaller stiffness than $\bar{\phi}_2$ ($K_1 < K_2$) and a larger relaxation rate and noise ($\Gamma_1 > \Gamma_2$). (We note that qualitatively similar results as the one presented in this section are obtained by considering conditions in which the asymmetry occurs in only one of the two parameters, i.e.,  $\Gamma_1 > \Gamma_2$ and $K_1 = K_2$ or $\Gamma_1 = \Gamma_2$ and $K_1 < K_2$.) When the feedback of the fluctuations is not taken into account, the order parameter again quickly recovers the initial value after the temperature quench  (orange dashed lines in Fig.~\ref{fig:rad_phi_vs_time_diff_GAM_AND_K_free_en}). Due to the inequivalent dynamics of the order parameters, there is some nontrivial evolution of the phase $\varphi$  (see also the plot of the trajectory in Fig.~\ref{fig:rad_phi_vs_time_diff_GAM_AND_K_free_en}(c)) when $R$ is small, but as soon as $R$ starts to recover its original magnitude the phase $\varphi$ collapses back to $\pi/4$.

We now take into account fluctuations. The behavior at early times, as long as only the weak initial equilibrium fluctuations are present, is again not affected by the fluctuations. Thereafter, as we observed also in the fully symmetric case discussed in the \textit{Isotropic fluctuations} section the system remains trapped in a state with $R \sim 0$ thanks to the renormalization of the quadratic coefficients of the free energy provided by the fluctuations. However, the trajectory of the recovery differs entirely from the symmetric case: Before the amplitude $R$ recovers the original magnitude $R \sim 4$, the system is transiently trapped in an intermediate state characterized by a plateau with $R\sim 3.85$ and $\varphi \sim \pi /2$ in the time window $t \sim 22-29$, see the rightmost part of Fig.~\ref{fig:rad_phi_vs_time_diff_GAM_AND_K_free_en}(a-b).

It is convenient to look at the free energy landscape renormalized by the fluctuations $\bar{F} (\bar{\phi}_1, \bar{\phi}_2)$ defined in Eq.~\eqref{eq:free_en_fluct} at several times during the dynamics of the order parameters to provide a more insightful interpretation of the formation of this state. In the time-window where the plateau is observed the unrenormalized free energy has long recovered the initial equilibrium form (see dashed orange line in Fig.~\ref{fig:r_n_vs_time_diff_GAM_AND_K}(a) for $r(T)$). Instead, the renormalized quadratic coefficients are substantially affected. In particular we observe $r^\text{fl}_{11} > r^\text{fl}_{22}$ as a consequence of a larger amount of fluctuations  in the $11$ channel compared to the $22$ channel ($n_{11} > n_{22}$), see the continuous lines in Fig.~\ref{fig:r_n_vs_time_diff_GAM_AND_K}(a-b). As a consequence, the free energy gets highly modified by the nonequilibrium onset of the anisotropic fluctuations, leading to the transient creation of a state which does not correspond to any minimum of the equilibrium free energy, see Fig.~\ref{fig:r_n_vs_time_diff_GAM_AND_K}(c-f). In particular, Fig.~\ref{fig:r_n_vs_time_diff_GAM_AND_K}(d) shows the shape of the free energy when the order parameter is trapped at $R \sim 0$, while Fig.~\ref{fig:r_n_vs_time_diff_GAM_AND_K}(e) shows $\bar{F} (\bar{\phi}_1, \bar{\phi}_2)$ when the order parameter is trapped at $R \sim 3.85$ and $\varphi \sim \pi/2$. Finally, Fig.~\ref{fig:r_n_vs_time_diff_GAM_AND_K}(f) shows an intermediate step before the order parameters recover the equilibrium value.

\begin{figure*}[tbp]
    \centerline{\includegraphics[width=0.98\textwidth]{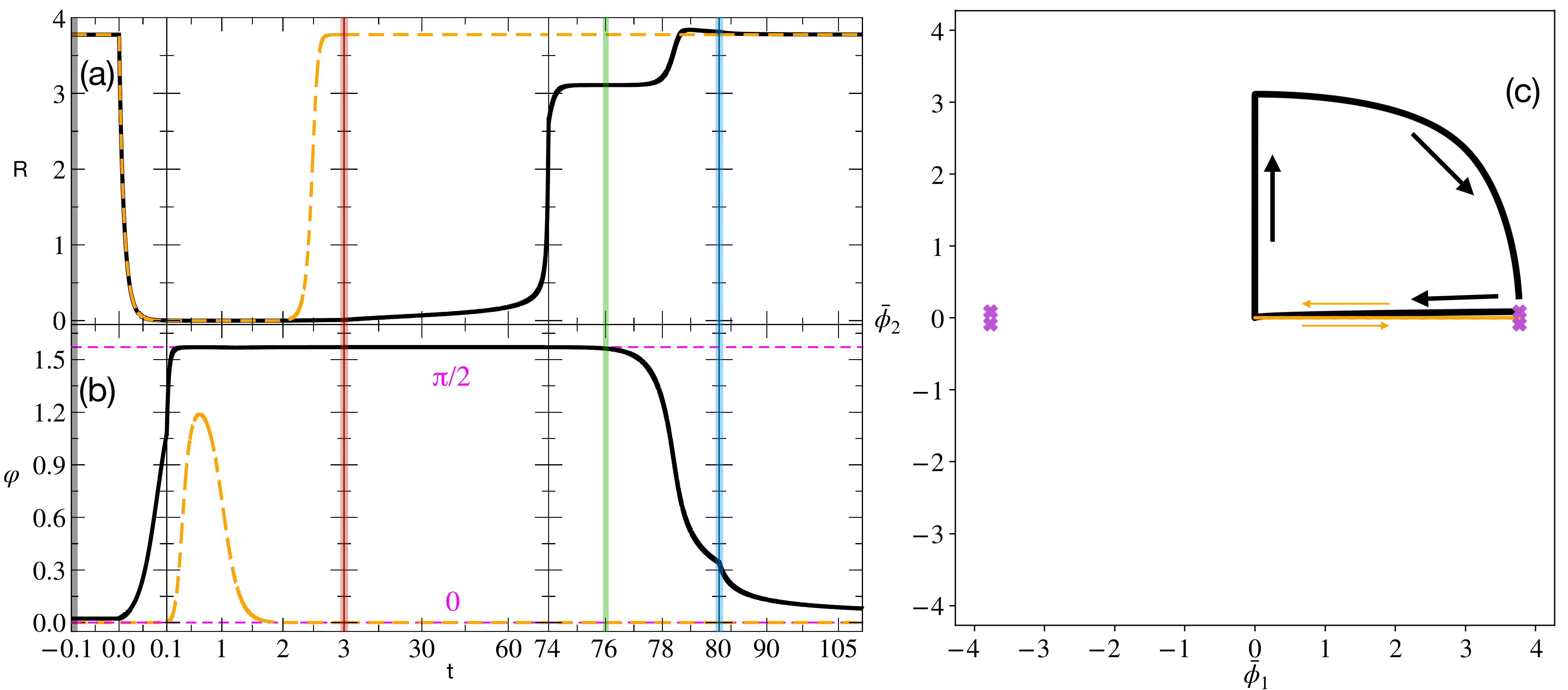}}
	\caption{ { \textbf{Amplitude, phase and trajectory for $\Gamma_1 > \Gamma_2$ and $K_1 < K_2$ after a temperature quench (almost competitive order) -} Dynamics of the amplitude $R$ (a) and the phase $\varphi$ (b) after a temperature quench (with the same properties described in the caption of Fig.~\ref{fig:rad_vs_time_sym}) for $\Gamma_1 = 0.9$, $\Gamma_2 = 0.5$, $K_1 = 0.1$ and $K_2 = 5.0$. The dashed orange lines correspond to the dynamics of $R$ and $\varphi$ without the feedback from the fluctuations. (c) Trajectories drawn by the order parameter in the $(\bar{\phi}_1, \bar{\phi}_2)$ plane during the time-dependent processes shown in panels (a-b). The light-violet crosses represent the position of the equilibrium minima in the renormalized potential. The parameters of the free energy are $T_\text{c} = 0.5$, $r_{1,0} = 15$, $r_{2,0} = 12$, $u_1 = 0.9$, $u_2 = 1.0$ and $\Lambda_\text{c} = 2 \pi$. This choice corresponds to the critical value $r_{2,0}/r_{1,0} = u_2/(2 u_1 - u_2)$, for which $\bar{\phi}_2 = 0$ in equilibrium in the absence of fluctuations. When fluctuations are included, $\bar{\phi}_2 \neq 0$ even if it remains small. The shaded vertical lines in panels (a-b) represent the times at which the renormalized free energies are displayed in Fig.~\ref{fig:r_n_vs_time_diff_GAM_AND_K_NOSYM}(c-f).}
    } \label{fig:rad_phi_vs_time_diff_GAM_AND_K_NOSYM_free_en}
\end{figure*}

The transient stabilization of the order $\bar \phi$ and the corresponding existence of a minimum of the renormalized free energy $\bar{F} (\bar{\phi}_1, \bar{\phi}_2)$ relies on the existence of an asymmetric fluctuation regime with $n_{11} > n_{22}$. An asymmetry $n_{11} > n_{22}$ is a priori expected because the fluctuations of $\phi_2$ decay faster than those of $\phi_1$ after the temperature quench due to the higher stiffness and the lower relaxation rate of the former with respect to the latter. However, we emphasize that this alone cannot explain a trapped regime in which $n_{22}$ and $n_{11}$ remain almost constant over a certain period of time. The latter is a consequence of the nonlinear behavior: While the stiffness sets the $k$-dependent energy cost for the fluctuations, the anharmonic interaction due to the quartic terms in $\mathcal{V}$ implies a mass term for the fluctuation spectrum which depends self-consistently on the value of the order parameter $\bar \phi$ (see the \textit{Methods} for the explicit equations). This mass term will itself be anisotropic if $\bar \phi \neq 0$. The trapping of the state at $\varphi=\pi/2$ is thus a consequence of mutual stabilization of nonthermal fluctuations and nonthermal order, i.e., of the NOBD mechanism.

\subsubsection*{(Almost) competitive order}
\begin{figure*}[tbp]
    \centerline{\includegraphics[width=0.98\textwidth]{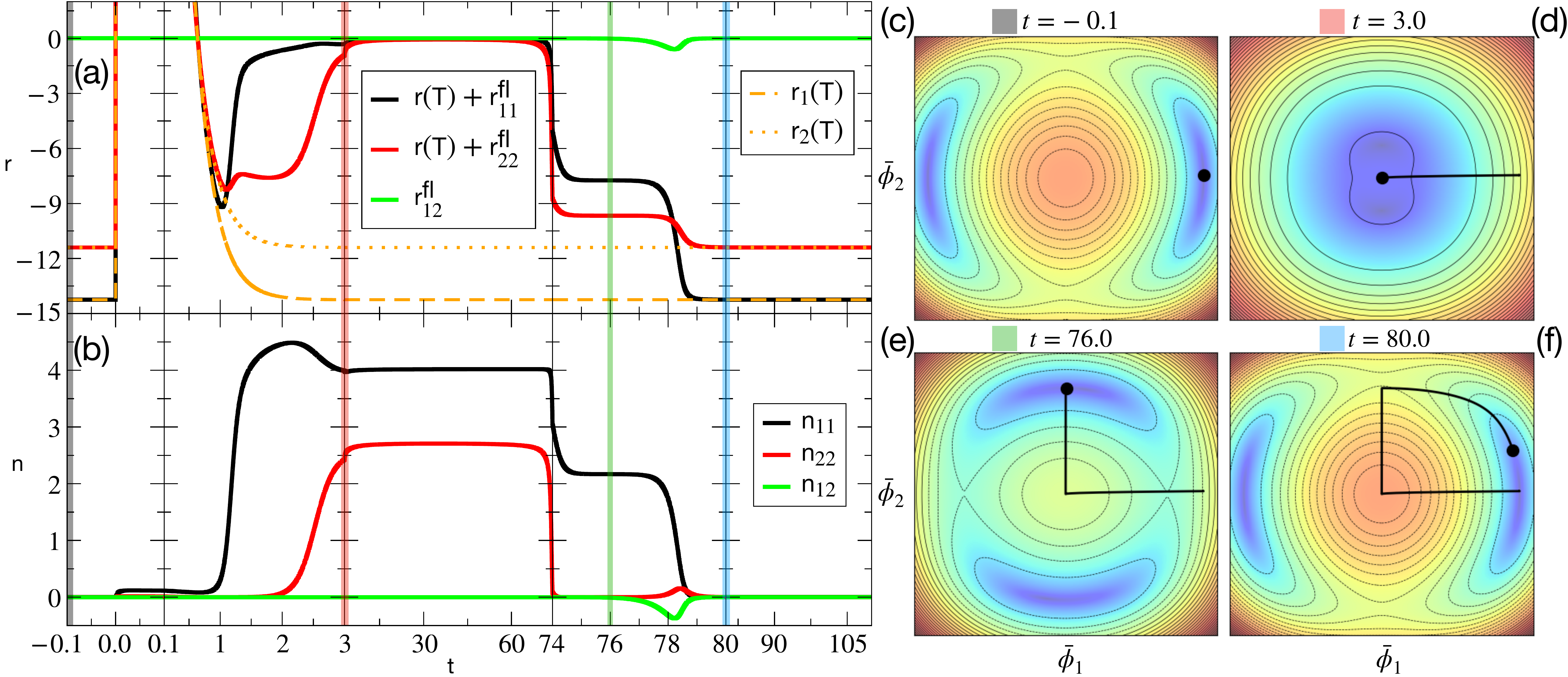}}
	\caption{ \textbf{Fluctuations and free-energy dynamics for $\Gamma_1 > \Gamma_2$ and $K_1 < K_2$ after a temperature quench (almost competitive order) -} Dynamics of the quadratic coefficients renormalized by the fluctuations (a) and of the excitations per channel (b) during the same time-dependent protocol considered in Figs.~\ref{fig:rad_vs_time_sym}-\ref{fig:rad_phi_vs_time_diff_GAM_AND_K_NOSYM_free_en}. The parameters are the ones described in the caption of Fig.~\ref{fig:rad_phi_vs_time_diff_GAM_AND_K_NOSYM_free_en}. The dashed (dotted) orange line in (a) represents the time-dependent quadratic coefficient $r_1 (T)$ ($r_2 (T)$) unrenormalized by the fluctuations. Panels (c-f) show the free energy renormalized by the fluctuations $\bar{F} (\bar{\phi}_1, \bar{\phi}_2 )$, see Eq.~\eqref{eq:free_en_fluct}, at several times before and after the temperature quench. The black dot shows the order parameter at a given time in the $(\bar{\phi}_1, \bar{\phi}_2 )$ plane, and the black line shows the trajectory drawn by the order parameters until the considered time. The shaded vertical lines in panels (a-b) represent the times at which the renormalized free energies are displayed. For the movie of the process, see Supplementary Note.}
    \label{fig:r_n_vs_time_diff_GAM_AND_K_NOSYM}
\end{figure*}

To discuss a case close to competitive order we consider the potential Eq.~\eqref{eq:pot_z2xz2_c4} with parameters $r_{1,0}\neq r_{2,0}$. More precisely, we take parameters corresponding to the critical ratio $r_{2,0}/r_{1,0} = u_2/(2 u_1 - u_2)$, beyond which $\bar{\phi}_2 = 0$ in equilibrium in the absence of fluctuations. The potential is again combined with the $\mathbb{Z}_2 \times \mathbb{Z}_2$ stiffness term Eq.~\eqref{eq:stiff_z2xz2} with $K_1 < K_2$, and we assume $\Gamma_1 > \Gamma_2$ as in the previous section. The combination of these two terms leads to the equilibrium free energy shown in Fig.~\ref{fig:r_n_vs_time_diff_GAM_AND_K_NOSYM}(c), which describes the case where one of the two order parameters is much larger than the other, $\bar{\phi}_1 \gg \bar{\phi}_2 \sim 0$. 

When fluctuations are neglected, the dynamics of the order parameters after a temperature quench is quite straightforward, see the dashed line in Fig.~\ref{fig:rad_phi_vs_time_diff_GAM_AND_K_NOSYM_free_en}(a). The amplitude $R$ collapses to $R \sim 0$, before recovering the original value. In the meantime, the phase $\varphi$ grows to a finite value, while $R \sim 0$, and then it goes back to $\varphi \sim 0$. We notice that this dynamic is possible because the phase in the equilibrium configuration is small but finite, $\varphi \sim 10^{-5}$ and $\Gamma_1 > \Gamma_2$. On practical grounds, the dynamics of the order parameter would not deviate from the line with $\bar{\phi}_2 = 0$ in this case, see Fig.~\ref{fig:rad_phi_vs_time_diff_GAM_AND_K_NOSYM_free_en}(c).

If, instead, the back action of the fluctuations is included, the time evolution becomes much different. In particular, before the system recovers the original amplitude $R$, a plateau at $R \sim 3$ and $\varphi \sim \pi/2$ appears in the time-window $t \sim 74-78$ indicative of the creation of a new configuration, see the black continuous lines in Fig.~\ref{fig:rad_phi_vs_time_diff_GAM_AND_K_NOSYM_free_en}(a-c). The dynamics can be interpreted in a similar way as in the previous section: The nonthermal fluctuations $n_{\alpha \beta}$ renormalize the quadratic coefficients (see Fig.~\ref{fig:r_n_vs_time_diff_GAM_AND_K_NOSYM}(a-b)) leading to the transient creation of a metastable state in correspondence of the plateau observed in the dynamics of $R$, see Fig.~\ref{fig:r_n_vs_time_diff_GAM_AND_K_NOSYM}(c-f).

\begin{figure*}[tbp]
    \centerline{\includegraphics[width=0.98\textwidth]{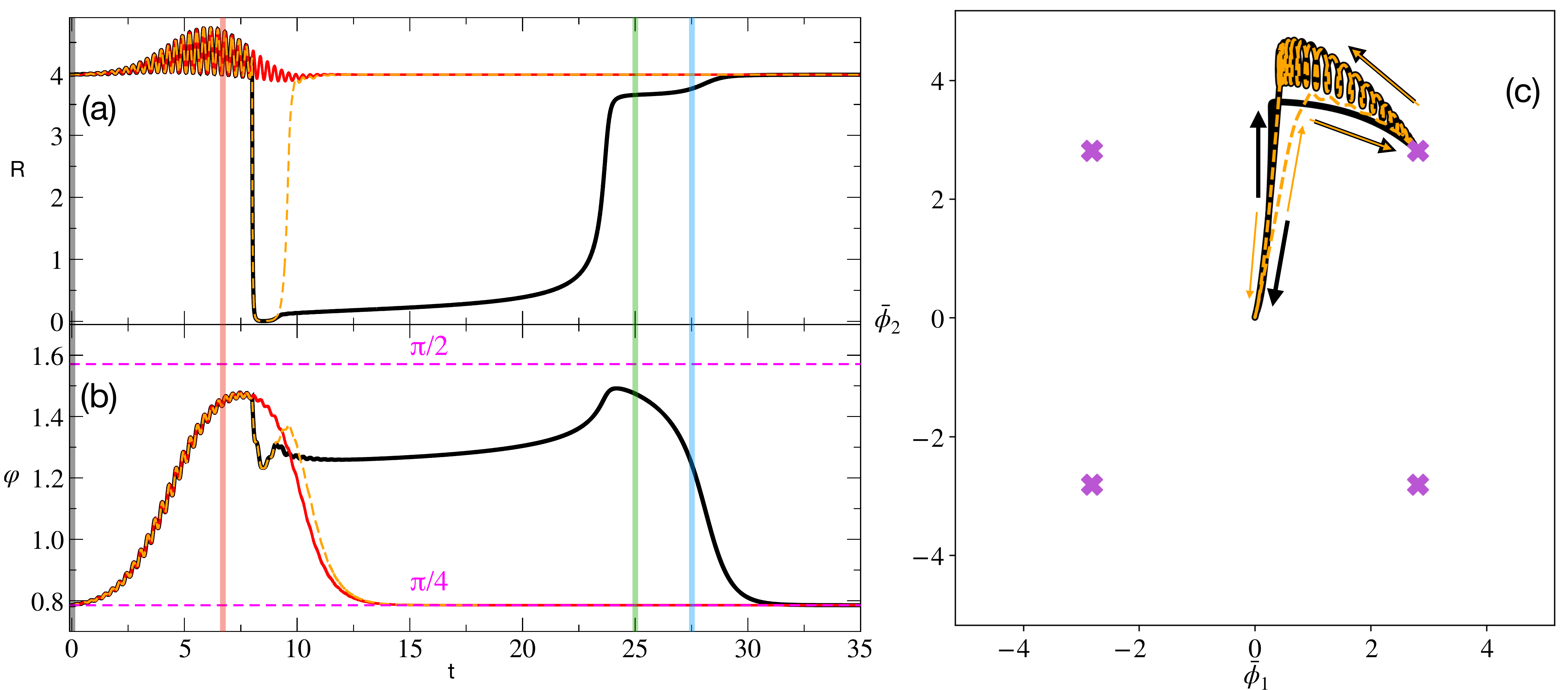}}
	\caption{ { \textbf{Amplitude, phase and trajectory for the $\text{C}_4$-symmetric model -} The black line shows the dynamics of the amplitude $R$ (a) and the phase $\varphi$ (b) after a coherent modulation of $J_2$ ($J_{1,0} = J_{2,0} = 7.5$, $\delta J_1 = 0$, $\delta J_2 = 5$, $\Omega = 10$, $t_\text{av} = 6$, $\sigma = 2$) and a temperature quench ($T_\text{i} = 0.025$, $T_\text{q} = 2.5$, $t_\text{q} = 8$, $\tau = 0.3$). The dashed orange lines correspond to the dynamics of $R$ and $\varphi$ without the feedback from the fluctuations. The solid red lines correspond to the dynamics in the presence of the fluctuations but without the temperature quench. (c) Trajectories drawn by the order parameter in the $(\bar{\phi}_1, \bar{\phi}_2)$ plane during the time-dependent processes shown in panels (a-b) (the red trajectory is not shown). The light-violet crosses represent the position of the equilibrium minima in the renormalized potential. The parameters of the free energy and for the dynamics are $T_\text{c} = 0.5$, $u_1 = 0.9$, $u_2 = 1.0$, $K = 5.0$, $\Gamma_1 = \Gamma_2 = 0.5$ and $\Lambda_\text{c} = \pi$. The shaded vertical lines in panels (a-b) represent the times at which the renormalized free energies are displayed in Fig.~\ref{fig:r_n_vs_time_c4sym}(c-f).}
    } \label{fig:rad_phi_vs_time_c4sym_free_en}
\end{figure*}

\subsection*{C$_4$ model}

In this section we demonstrate the mechanism of NOBD in yet another interesting situation. The model represents a situation in which the system has a higher symmetry in equilibrium, which is then transiently lowered during the pulse. Fluctuations are then found to stabilize a nonthermal order which reflects the lower symmetry, even long times after the perturbation that has explicitly broken the symmetry.

Specifically, we analyze the case in which the free energy Eq.~\eqref{eq:free_en} has $\text{C}_4$ symmetry in equilibrium, i.e., the interaction potential has the form shown in Eq.~\eqref{eq:pot_z2xz2_c4} with $r_{\alpha, 0} \equiv 2 J_\alpha$ and $J_1 = J_2$, while the stiffness contribution Eq.~\eqref{eq:stiff_gen} has coefficients $\bar{K}_{i, \alpha} = r_{\alpha, 0} K \delta_{i,\alpha}$, leading to:
\begin{align} \label{eq:stiff_c4}
    \mathcal{K} (\phi_1, \phi_2) = \frac{K}{2} \sum_\alpha r_{\alpha, 0} (\partial_\alpha \phi_\alpha)^2 .
\end{align}
In Eq.~\eqref{eq:stiff_c4}, $\partial_\alpha$ indicates the spatial derivative along the $\alpha$ direction, which makes the fluctuation contribution isotropic in the two order parameters even if the problem becomes spatially anisotropic (we consider the model in spacial dimension $d=2$). The parameters of the potential Eq.~\eqref{eq:pot_z2xz2_c4} will be chosen such that the bare free energy (without fluctuations) is the same as considered in the \textit{Cooperative order} section.

With the stiffness term \eqref{eq:stiff_c4}, the model is invariant under the transformation $(\phi_1 , \phi_2 ; k_1 , k_2) \rightarrow (\phi_2 , -\phi_1 ; k_2 , -k_1)$. When we talk about the $\text{C}_4$ symmetry of the model, we refer to this transformation. The model is equivalent to the classical and continuum limit of the $90^\circ$ compass model \cite{Nussinov2015_RMP} used to describe orbital ordered systems \cite{Mishra2004_PRL} and $p + i p$ SC \cite{Nussinov2005_PRB}. Within the correspondence with the orbital ordered systems, the $J_\alpha$ coefficients represent the coarse-grained superexchange interactions $\bar{J}_\alpha$ responsible for the orbital order in the microscopic model, see the \textit{Methods}. Thus, it is possible to time-dependently control $J_\alpha$ by means of Floquet-driving \cite{Mentink2015_NatComm,Itin2015_PRL,Bukov2015_AdvPhys,Eckstein2017_arXiv,Grandi2021_PRB,Mueller2022_PRB}, by using suitable linearly polarized pump-pulses or by launching a ps current pulse along a specific direction of the sample. All these protocols, and others, can transiently break the symmetry between the two order parameters.

\begin{figure*}[tbp]
    \centerline{\includegraphics[width=0.98\textwidth]{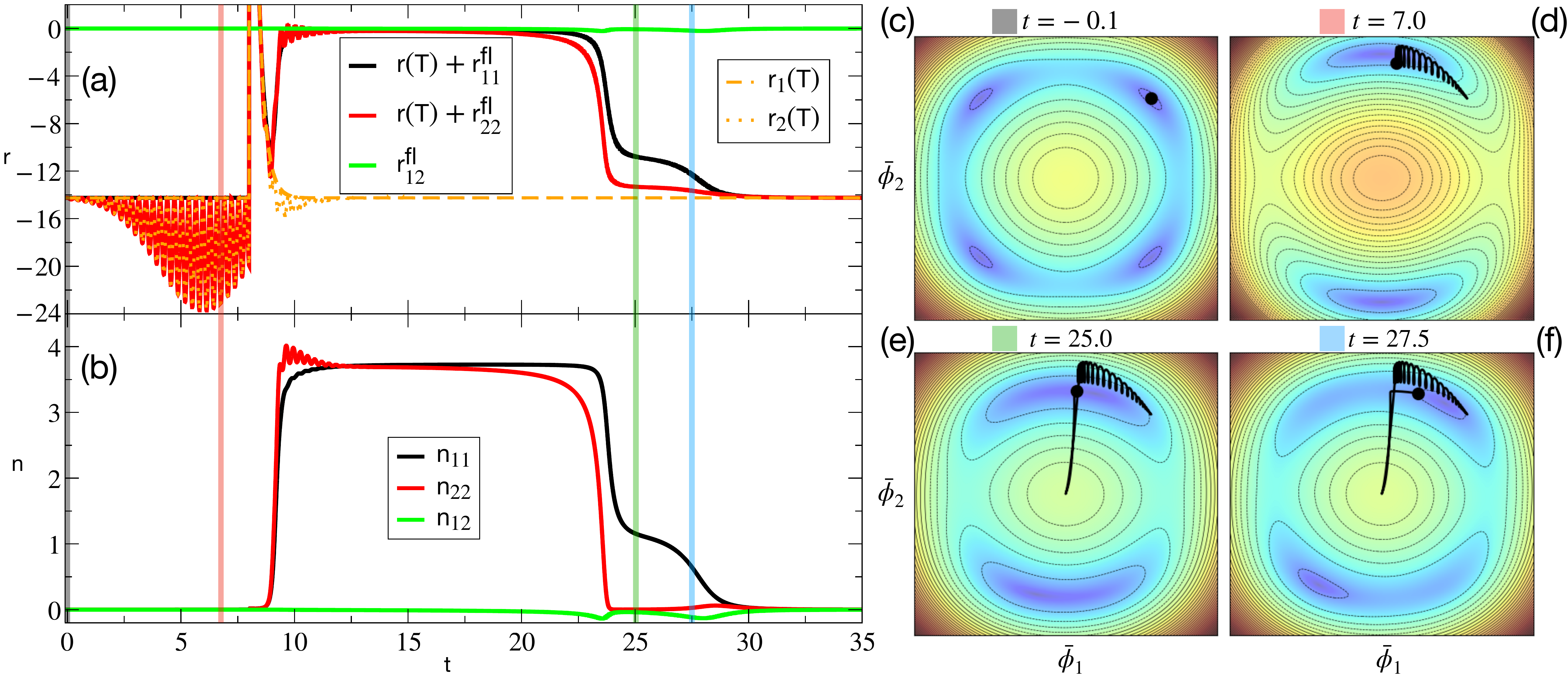}}
	\caption{ \textbf{Fluctuations and free-energy dynamics for the $\text{C}_4$-symmetric model -} Dynamics of the quadratic coefficients renormalized by the fluctuations (a) and of the excitations per channel (b) during the same time-dependent protocol considered in Fig.~\ref{fig:rad_phi_vs_time_c4sym_free_en} (coherent control of $J_2$ plus incoherent temperature quench). The dashed (dotted) orange line in (a) represents the unrenormalized time-dependent quadratic coefficient $r_1 (T)$ ($r_2 (T)$). Panels (c-f) show the free energy renormalized by the fluctuations $\bar{F} (\bar{\phi}_1, \bar{\phi}_2 )$, see Eq.~\eqref{eq:free_en_fluct}, at several times before and after the temperature quench. The black dot shows the order parameter at a given time in the $(\bar{\phi}_1, \bar{\phi}_2 )$ plane, and the black line shows the trajectory drawn by the order parameters until the considered time. The shaded vertical lines in panels (a-b) represent the times at which the renormalized free energies are displayed. For the movie of the process, see the Supplementary Note.}
    \label{fig:r_n_vs_time_c4sym}
\end{figure*}

\begin{figure*}[tbp]
    \centerline{\includegraphics[width=1.0\textwidth]{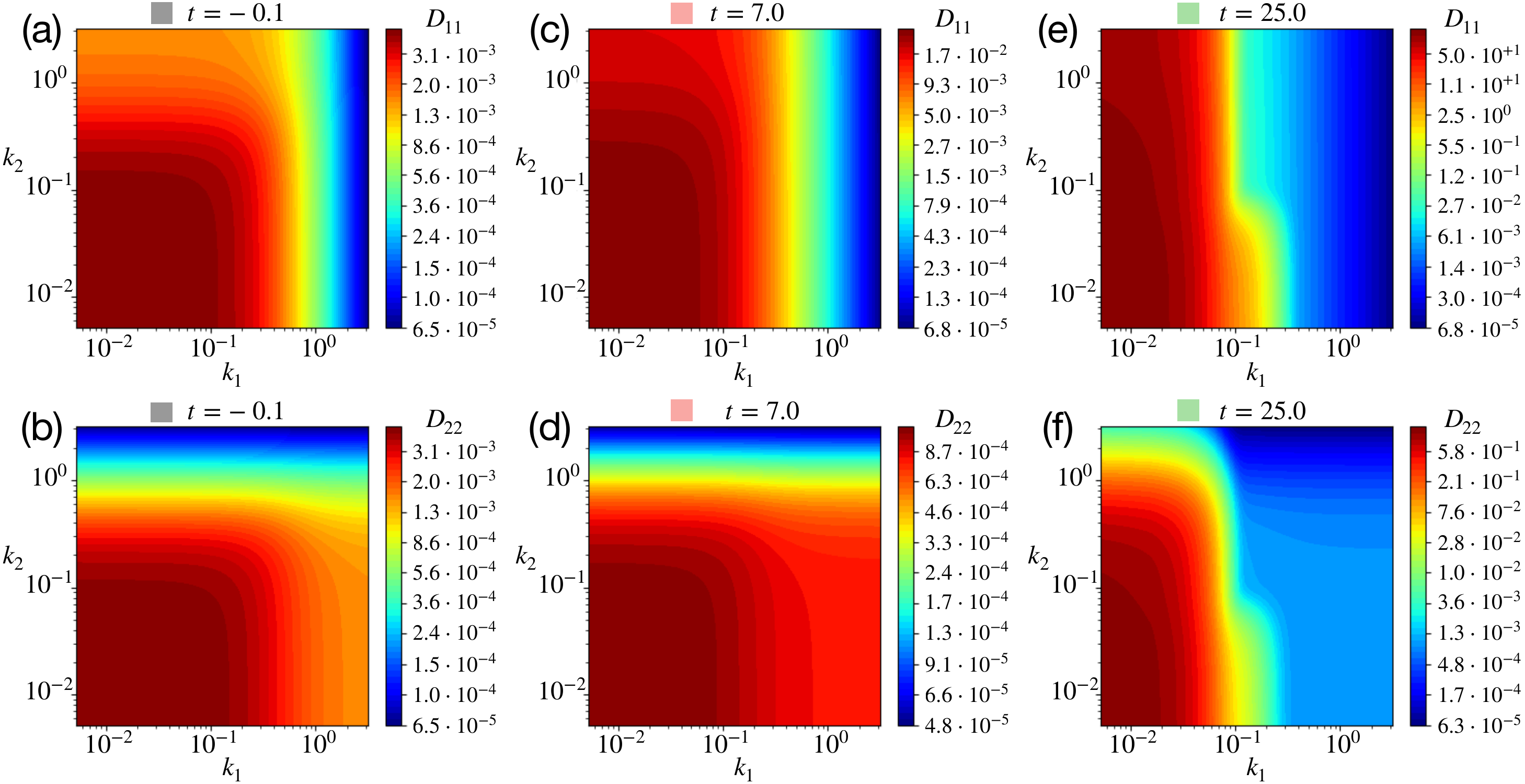}}
	\caption{ \textbf{$k$-resolved time-dependent distribution of the fluctuations -} Momentum-resolved diagonal correlation functions at several time-steps (a and b at $t=-0.1$, c and d at $t=7.0$, e and f at $t=25.0$) during the time-dependent excitation discussed in the caption of Fig.~\ref{fig:rad_phi_vs_time_c4sym_free_en}. The axis $k_1$ and $k_2$ are in logarithmic scale. The upper (lower) row shows $D_{11}$ ($D_{22}$).}
    \label{fig:corr_funct}
\end{figure*}

In particular, we consider a time-dependent modulation of the kind:
\begin{align} \label{eq:J_time_dep}
    J_\alpha (t) = J_{\alpha, 0} + \delta J_\alpha \sin^2 (\Omega t) \ s (t) ,
\end{align}
where $J_{\alpha, 0}$ is the equilibrium value of the coarse-grained superexchange, $\delta J_\alpha$ is the maximum change of $J_{\alpha, 0}$, and $\Omega$ is the frequency of, e.g., an electric field acting in the $\alpha$ direction \cite{Eckstein2017_arXiv}. The envelope function $s(t) = \exp \big[- (t - t_\text{av})^2/(2 \sigma^2) \big] \theta (t)$ is a Gaussian centered at the time $t_\text{av}$ and with variance $\sigma$.

Within this section, we assume $\Gamma_1 = \Gamma_2$, meaning that the symmetry among the fields $\bar{\phi}_1$ and $\bar{\phi}_2$ is also preserved dynamically. However, the specific time-dependent protocol considered to excite the system can transiently break the $\text{C}_4$ symmetry in one of the ways discussed above (Floquet driving, linearly polarized pump-pulses or ps current pulses along a specific direction of the sample). The dynamics of the order parameters induced by the excitation protocol Eq.~\eqref{eq:J_time_dep} is shown by the red curves in Figs.~\ref{fig:rad_phi_vs_time_c4sym_free_en}(a-b), which describe the time evolution of $R$ and $\varphi$ under the action of the time-modulation of the $J_2$ interaction. In this case, the amplitude $R$ oscillates in the proximity of the original value, and the phase $\varphi$ moves from $\pi/4$ to $\sim \pi/2$ and then goes back to the equilibrium configuration when the pulse is over (roughly at $t \sim 12$).

While the coherent driving transiently breaks the $\text{C}_4$ symmetry of the problem, one can apply a second incoherent pulse (temperature quench), that has the effect of suppressing the amplitude $R$ and of increasing the magnitude of the fluctuations, as shown by the black line in Figs.~\ref{fig:rad_phi_vs_time_c4sym_free_en}(a) and by Fig.~\ref{fig:r_n_vs_time_c4sym}(b), respectively. When $R$ starts to depart from the region near zero, $\varphi$ approaches once again $\pi/2$, suggesting, even in this case, the creation of a new transient configuration in a region where the equilibrium free energy does not show any minima. At the time in which the metastable state is observed (corresponding to the emergence of a plateau in the time-evolution of $R$), the time-modulation of $J_2$ is over and thus cannot be responsible for the observed behavior. Instead, the fluctuations play the most relevant role in determining the transition to this metastable state, as evident by comparing the dynamics when fluctuations are disregarded (orange lines in Fig.~\ref{fig:rad_phi_vs_time_c4sym_free_en}(a-c)) with the dynamics with the feedback of the fluctuations (black lines in Fig.~\ref{fig:rad_phi_vs_time_c4sym_free_en}(a-c)).

The fluctuations lead indeed to a renormalization of the bare quadratic coefficients, compare the dashed and the dotted lines with the solid lines in Fig.~\ref{fig:r_n_vs_time_c4sym}(a), which produce a change in the shape of the free energy of the system, as displayed in Fig.~\ref{fig:r_n_vs_time_c4sym}(c-f). As shown by Fig.~\ref{fig:r_n_vs_time_c4sym}(e), a new state gets transiently created even in this configuration.

Since the stiffness contribution to the free energy is spatially anisotropic, see Eq.~\eqref{eq:stiff_c4}, one can get the most complete characterization of the dynamics of the system by inspecting the time-evolution of the $k$-distribution of the correlation matrix $D(\mathbf{k},t)$. To simplify the discussion, in Fig.~\ref{fig:corr_funct} we focus on the $k$-resolved diagonal elements of $D(\mathbf{k},t)$ at several times during the dynamics displayed in Figs.~\ref{fig:rad_phi_vs_time_c4sym_free_en}-\ref{fig:r_n_vs_time_c4sym}. As compared to the equilibrium distribution displayed in Fig.~\ref{fig:corr_funct}(a-b), the nonequilibrium ones are remarkably different both in terms of shapes and magnitudes. In equilibrium, the correlation functions have the symmetry $D_{11} (k_1, k_2) = D_{22} (k_2, k_1)$, reflecting the $\text{C}_4$ symmetry of the model. Out of equilibrium, this symmetry is lost. At time $t = 7.0$, the coherent pulse is still active while the incoherent excitation has yet to arrive. At this time, the amplitude of the order parameter $R$ is large, and the phase of the system is $\varphi \sim \pi/2$, see Fig.~\ref{fig:rad_phi_vs_time_c4sym_free_en}(a-b). The $k$-distribution of the correlation functions is not very different from the equilibrium case in terms of magnitude (see the color bar of Fig.~\ref{fig:corr_funct}(c-d)). However, the shape changes, resulting in an elongation along the $k_2$ direction which is particularly visible in $D_{11} (\mathbf{k},t)$, see Fig.~\ref{fig:corr_funct}(c). When $t=25.0$, the coherent driving has already ended, and the effect of the temperature quench is still sizable; the amplitude of the order parameter $R$ and the phase $\varphi$ are similar to the ones at $t=7.0$, see Fig.~\ref{fig:rad_phi_vs_time_c4sym_free_en}(a-b), however, the distribution of the fluctuations is much different at the two times, compare Fig.~\ref{fig:corr_funct}(c-d) and Fig.~\ref{fig:corr_funct}(e-f). Indeed, at $t=25.0$ both $D_{11} (\mathbf{k},t)$ and $D_{22} (\mathbf{k},t)$ are highly elongated along $k_2$, and the magnitude of the correlation functions is strongly enhanced with respect to the values it had before the incoherent excitation, compare the color bar of Fig.~\ref{fig:corr_funct}(e-f) with the ones of Fig.~\ref{fig:corr_funct}(a-d). After this time, the system goes back to the starting equilibrium configuration, and the distribution of the fluctuations becomes again the one shown in Fig.~\ref{fig:corr_funct}(a-b).

The spatially anisotropic character of the nonequilibrium fluctuations shown in Fig.~\ref{fig:corr_funct} might be probed by means of time-resolved elastic techniques such as resonant elastic x-ray scattering \cite{Fink2013_RPP}. Another possibility to explicitly probe the time-evolution of the transient disorder is given by ultrafast transmission electron microscopy \cite{Domroese2023_NatMat}.

\section*{Discussion}
In the previous sections, we have analyzed the effect of the stiffness anisotropies on the nonequilibrium dynamics in two models described by two order parameters $\bar{\phi}_1$ and $\bar{\phi}_2$ governed by $\mathbb{Z}_2 \times \mathbb{Z}_2$ and $\text{C}_4$ symmetry, respectively. The former model represents a minimal (toy-model) description of the competition or the cooperation between two generic ordered states, while the latter describes a toy-model for orbitally ordered systems. We show that a nonequilibrium excitation can lead to the transient stabilization of a state that is not a minimum of the equilibrium free energy by the nonthermal fluctuations of the order parameters. In the former case, this effect can be achieved by a single incoherent excitation, i.e., a temperature quench, while in the latter, it is crucial to first reduce the $\text{C}_4$ symmetry of the problem through the excitation. Our work represents a \textit{proof of principle} of the shaping of the free energy landscape of a system by nonequilibrium fluctuations, a phenomenon that might be regarded as a generalization of the order-by-disorder mechanism to the nonequilibrium realm (NOBD). The observation of this effect already in the minimal models described above suggests the potential generality of the mechanism.

We now focus on a few possible specific physical interpretations of the order parameters $\bar{\phi}_1$ and $\bar{\phi}_2$. The first interpretation we examine involves the coexistence of density wave and SC as ordered states, a phenomenon observed in several material classes, such as cuprates, iron-based superconductors, and vanadium-based kagome metals. Certain cuprate compounds may be relevant for the NOBD mechanism described in this paper. For instance, La$_{2-x}$Ba$_{x}$CuO$_4$ ($x = 0.115$) shows coexistent charge (T$_\text{co} \sim 53$K), spin (T$_\text{so} \sim 40$K) and superconducting (T$_\text{c} \sim 13$K) orders. Below T$_\text{c}$, the system can be converted into a strong-superconducting phase by means of a near-infrared pump pulse with fluence $\sim 0.1$mJ$/$cm$^2$. The effect disappears if the equilibrium temperature is even slightly above T$_\text{c}$ \cite{Cremin2019_PNAS}. This condition is similar to the one analyzed for the $\mathbb{Z}_2 \times \mathbb{Z}_2$ model, where the transient stabilization of the superconducting state (the order parameter $\bar{\phi}_2$ in our notation) might be induced by a different stiffness of the two order parameters, by different relaxation rates or by a combination of both. For the superconducting state to dominate out of equilibrium, a non-zero value of $\bar{\phi}_2$, even if small, must be present in the equilibrium configuration in our toy-model description. Conversely, the photoexcitation of YBa$_2$Cu$_3$O$_{6+x}$ ($x=0.67$) in the regime where a charge density wave (CDW) is found to coexist with SC leads to a transient increase of CDW at the expense of SC \cite{Wandel2022_Science}. For this species of curates, the ratio of stiffness and relaxation rate of CDW and SC behaves oppositely to La$_{2-x}$Ba$_{x}$CuO$_4$.

Another intriguing class of materials that show a coexistence, in equilibrium, of SC and CDWs, is given by the vanadium-based kagome metals. Even if, at present, no experimental evidence is currently available regarding the photoinduced enhancement of SC or density waves in these compounds, we envision them to be a very promising setup for the investigation of NOBD. Indeed, there is a growing experimental \cite{Chen2022_PRL,Yang2023_PRB,Sur2023_NatComm,Subires2023_NatComm} and theoretical \cite{Tazai2022_SciAdv,Tsvelik2023_PRB,Tazai2023_NatComm,Grandi2024_PRB,Tian2024_arXiv} evidence of the important role played by fluctuations in the kagome metals in equilibrium, rendering them promising for revealing nonequilibrium phenomena.

$\bar{\phi}_1$ and $\bar{\phi}_2$ may also represent different orbital order configurations. Given the directional nature of the orbital degree of freedom, real-space anisotropies are well expected for systems that describe orbital-orbital interactions. This might be a representative case analyzed above through the C$_4$-symmetric model. Our theory might provide a different (phenomenological) explanation of light-induced hidden states in spin and orbital-ordered systems \cite{Li2018_NatComm}. From an experimental perspective, our theory might provide an explanation of the nonthermal orbital order transiently induced in Nd$_{0.5}$Sr$_{0.5}$MnO$_3$ \cite{Ichikawa2011_NatMat} via the NOBD mechanism.

A combination of the two previously-mentioned scenarios, i.e., direct stiffness and interaction potential anisotropy in the order parameters $\bar{\phi}_1$ and $\bar{\phi}_2$ and spatial anisotropy of the stiffness might be representative of the physics of vanadium dioxide, for which a $\mathbb{Z}_2 \times \mathbb{Z}_2$ symmetric free energy has been previously identified \cite{Grandi2020_PRR}. In that case, the two order parameters describe the two components of the lattice distortion that take place in the system when moving from the high-temperature metallic rutile phase to the low-temperature insulating monoclinic state (with transition temperature T$_\text{c} \sim 340$K) \cite{Goodenough1971_JSolStChem}. The first component of the structural change is a dimerization involving the vanadium atoms that takes place along the rutile c-axis, while the second is a tilting component of the vanadium dimers out of the c-axis. By their own nature, the dimerization is an almost one-dimensional order parameter, while the tilting is mainly two-dimensional (which already provides a spatial anisotropy that also affects the respective stiffness). Moreover, the magnitude of the stiffness constants for dimerization and tilting must differ due to their distinct couplings with the electronic degree of freedom, which is integrated out in the effective theory. Specifically, the tilting lifts the band degeneracy among the three t$_{2 \text{g}}$ orbitals, leaving only one of them at the Fermi level. In contrast, the dimerization opens an energy gap in the left-alone band. Whether it is possible to stabilize by such a mechanism a metallic monoclinic state in VO$_2$ in which just one of the two structural distortions is active remains to be understood, but previous experimental evidence suggests the presence of two time scales in the \textit{melting} of the monoclinic structure after photoexcitation \cite{Baum2007_Science}, in analogy with what we have discussed in the previous sections. The main limitation of the comparison between the present study and the physics of VO$_2$ comes from the fact that, in vanadium dioxide, the order parameters have a structural nature, i.e., the dynamics might be affected by the coherent oscillations coming from a mass term proportional to a second order time derivative. Exploring this additional feature goes beyond the scope of the present work.

\section*{Methods}

\subsection*{Full dynamical equations}  \label{app_a}
We report here the full dynamical equations of the problem. The momentum $\mathbf{q}$-dependent average order parameter is defined as:
\begin{align}
    \bar{\phi}_{\alpha, \mathbf{q}} (t) = \int \mathcal{D} [\phi] \phi_{\alpha, \mathbf{q}} \mathcal{P} [\phi] (t) , \nonumber 
\end{align}
with $\mathcal{P} [\phi] (t)$ the probability distribution functional of the field configurations $\phi_{\alpha, \mathbf{q}}$ at time $t$. The average field considered in the main text is the homogeneous ($\mathbf{q} = \mathbf{0}$) component of the field just defined $\bar{\phi}_{\alpha} (t) = \bar{\phi}_{\alpha, \mathbf{q} = \mathbf{0}} (t)$ \cite{Mazenko1985_PRB,Zannetti1993_JPA}. From a real space perspective, this average field corresponds to the average over the field realizations as well as over space. Considering the homogeneous contribution as the most relevant one is justified when spatially homogeneous perturbations are taken into account, as it is always the case in this work. The time evolution of the homogeneous average field $\bar{\phi}_{\alpha} (t)$ is given by:
\begin{align}
    \partial_t \bar{\phi}_{\alpha} (t) = - \Gamma_\alpha \langle \frac{\delta \mathcal{F}}{\delta \phi_{\alpha, \mathbf{0}}} \rangle , \nonumber
\end{align}
where the functional derivative of the free energy is computed with respect to the zero-momentum (homogeneous) field $\phi_{\alpha, \mathbf{0}}$ and the average is taken with respect to the probability distribution $\mathcal{P} [\phi] (t)$. By assuming $\mathcal{P} [\phi] (t)$ to be Gaussian around the homogeneous average field at time $t$, which corresponds to the Gaussian approximation, we can apply Wick's theorem for the evaluation of $\langle \delta \mathcal{F} / \delta \phi_{\alpha, \mathbf{0}} \rangle$, leading to:
\begin{align}
    \langle \frac{\delta \mathcal{F}}{\delta \phi_{\alpha, \mathbf{0}}} \rangle \approx \sum_{\beta = 1}^2 r^\text{eff}_{\alpha \beta} \bar{\phi}_\beta (t) , \nonumber
\end{align}
where:
\begin{align}
    & r^\text{eff}_{\alpha \beta} = \big[ r_\alpha (T) + u_2 (\bar{\phi}_\alpha^2 + 3 n_{\alpha \alpha}) \nonumber  \\
    & + (2 u_1 - u_2) (\bar{\phi}_{\bar{\alpha}}^2 + n_{\bar{\alpha} \bar{\alpha}} - 2 n_{\alpha \bar{\alpha}}) \big] \delta_{\alpha \beta} \nonumber \\
    & + 2 (2 u_1 - u_2) n_{\alpha \beta} \nonumber
\end{align}
is the effective quadratic coefficient that couples the average fields $\bar{\phi}_\alpha$ and $\bar{\phi}_\beta$ renormalized by the fluctuations $n_{\alpha \beta}$, see Eq.~\eqref{eq:fluct}. Even if every single pump-probe experiment might show a behavior different from the one described by the dynamics of the average $\bar{\phi}_\alpha$, still each of them shows a tendency towards the predicted dynamics given the Gaussian probability distribution $\mathcal{P} [\phi] (t)$ we assumed. The equations of motion for $\bar{\phi}_{\alpha} (t)$ are supplemented by the equation of motion for the correlation functions at each k-point, which are computed using the same approximations discussed above:
\begin{align} \label{eq_app:dyn_corr_funct}
    \partial_t D_{\alpha \beta} (\mathbf{k}, t) & = \big( \Gamma_\alpha + \Gamma_\beta \big) T \delta_{\alpha \beta} \nonumber \\
    & - \Gamma_\alpha \sum_\gamma M_{\alpha \gamma} (\mathbf{k}, t) D_{\gamma \beta} (\mathbf{k}, t) \nonumber \\
    & - \Gamma_\beta \sum_\gamma M_{\beta \gamma} (\mathbf{k}, t) D_{\gamma \alpha} (\mathbf{k}, t) , 
\end{align}
with:
\begin{align}
    & M_{\alpha \beta} (\mathbf{k}, t) = r^\text{eff}_{\alpha \beta} + 2 (2 u_1 - u_2) \bar{\phi}_\alpha \bar{\phi}_\beta \nonumber \\
    & + \delta_{\alpha \beta} \big[ \sum_i \tilde{K}_{\alpha, i} \ k_i^2 + 2 u_2 \bar{\phi}_\alpha^2 - 2 (2 u_1 - u_2) \bar{\phi}_\alpha \bar{\phi}_\beta \big] . \nonumber
\end{align}
The time-dependent equations above are derived assuming a Gaussian probability distribution, i.e., Wick's theorem can be applied. Differently from other approaches \cite{Dolgirev2020_PRB}, the large-N expansion is not performed given that we have two order parameters ($N=2$) and the analyzed potentials do not have a continuous symmetry \cite{Brihaye1985_PLB}.
In equilibrium, Eqs.~\eqref{eq_app:dyn_corr_funct} can be analytically solved leading to the expression for the correlation functions:
\begin{align}
    D_{\alpha \beta} (\mathbf{k}) = (-1)^{\alpha + \beta} \frac{T \ M_{\bar{\alpha} \bar{\beta}} (\mathbf{k})}{\det \big[ M (\mathbf{k}) \big]} , \nonumber
\end{align}
where $\det \big[ M (\mathbf{k}) \big] = M_{1 1} (\mathbf{k}) M_{2 2} (\mathbf{k}) - M_{1 2} (\mathbf{k})^2$.

\subsection*{$90^\circ$ compass model}  \label{app_b}
The $90^\circ$ compass model on the square lattice is written as \cite{Mishra2004_PRL,Nussinov2005_PRB,Nussinov2015_RMP}
\begin{align} \label{eq_app:ham_compass}
    H = - \sum_{\mathbf{r}, \alpha} J_\alpha \tau_{\mathbf{r}, \alpha} \cdot \tau_{\mathbf{r} + \mathbf{e}_\alpha, \alpha} ,
\end{align}
where $\mathbf{r}$ runs over the lattice sites and $\boldsymbol{\tau}_{\mathbf{r}} = (\tau_{\mathbf{r},1}, \tau_{\mathbf{r},2})$ is a pseudospin vector where the components are the first two Pauli matrices. The interaction is spatially anisotropic and not SU$(2)$-invariant since just the pseudospin component $\tau_{\alpha}$, with $\alpha = 1,2$, is involved in the interaction along the spatial direction $\mathbf{e}_\alpha$ ($\mathbf{e}_1 = (1,0)$ or $\mathbf{e}_2 = (0,1)$). When $J_1 = J_2$, the model is invariant under the action of a global fourfold rotation which simultaneously rotates the lattice and the pseudospins by $n \pi /4$, with $n \in \mathbb{Z}$. Moreover,  Eq.~\eqref{eq_app:ham_compass} is invariant under the transformation $(\tau_{\mathbf{r}, 1}, \tau_{\mathbf{r}, 2}) \rightarrow (-\tau_{\mathbf{r}, 1}, \tau_{\mathbf{r}, 2})$ ($(\tau_{\mathbf{r}, 1}, \tau_{\mathbf{r}, 2}) \rightarrow (\tau_{\mathbf{r}, 1}, -\tau_{\mathbf{r}, 2})$) for $\mathbf{r}$ belonging to a given row (column). When $J_1 \neq J_2$, the first of the two symmetries gets broken down to $\mathbb{Z}_2 \times \mathbb{Z}_2$.

\noindent
One can derive the continuum and classical limit of Eq.~\eqref{eq_app:ham_compass} by rewriting the previous Hamiltonian as:
\begin{align}
    H & = - \sum_{\mathbf{r}, \alpha} J_\alpha \tau_{\mathbf{r}, \alpha} \cdot \tau_{\mathbf{r} + \mathbf{e}_\alpha, \alpha} \pm \sum_{\mathbf{r}, \alpha} J_\alpha \tau_{\mathbf{r}, \alpha} \tau_{\mathbf{r}, \alpha} \nonumber \\
    & = \frac{1}{2} \sum_{\mathbf{r}, \alpha} J_\alpha \big( \tau_{\mathbf{r}, \alpha} - \tau_{\mathbf{r} + \mathbf{e}_\alpha, \alpha} \big)^2 - \sum_{\mathbf{r}, \alpha} J_\alpha \tau_{\mathbf{r}, \alpha} \tau_{\mathbf{r}, \alpha} \nonumber \\
    & = \frac{1}{2} \sum_{\mathbf{k}, \alpha} J_\alpha E_\alpha (\mathbf{k} \cdot \mathbf{e}_\alpha ) \tau_{\mathbf{k}, \alpha} \tau_{-\mathbf{k}, \alpha} - \sum_{\mathbf{k}, \alpha} J_\alpha \tau_{\mathbf{k}, \alpha} \tau_{-\mathbf{k}, \alpha} , \nonumber
\end{align}
where in the last step we have performed a Fourier transformation and introduced the k-vector $\mathbf{k} = (k_1, k_2)$ and the dispersion $E_\alpha (\mathbf{k} \cdot \mathbf{e}_\alpha ) = 2 [ 1 - \cos (\mathbf{k} \cdot \mathbf{e}_\alpha) ]$. In the continuum limit, $E_\alpha (\mathbf{k} \cdot \mathbf{e}_\alpha ) \approx (\mathbf{k} \cdot \mathbf{e}_\alpha)^2$, and by mapping the pseudospins into classical vectors and transforming back to real space, one arrives at the quadratic part of the C$_4$-symmetric model introduced in the \textit{Results} section, see Eq.~\eqref{eq:pot_z2xz2_c4} and Eq.~\eqref{eq:stiff_c4}, where the quadratic coefficeints are made temperature dependent in the Ginzburg-Landau approach. The quartic terms have the role of keeping the length of the order parameters finite, and they are the ones compatible with the symmetry of the model when $J_1 = J_2$, which is the equilibrium configuration we start from.

\section*{Resources Availability}

\subsection*{Lead contact}
Further information and requests should be directed to and will be fulfilled by Francesco Grandi (francesco.grandi.89@gmail.com).

\subsection*{Materials availability}
This study did not generate new materials.

\subsection*{Data and code availability}
The data generated by the present study that are used in the figures of the Main Manuscript (Figs. 2 to 10) and the videos of Supplementary Information can be made available after request to the corresponding author.

\section*{Acknowledgment}
F.G. acknowledges discussions with Aaron M\"{u}ller on the Gaussian approximation and the large-N expansion. F.G., A.P. and M.E. acknowledge stimulating discussions on topics related to the present work with Simon Wall. F.G. and D.M.K. acknowledge support by the Deutsche Forschungsgemeinschaft (DFG, German Research Foundation) via Germany’s Excellence Strategy-Cluster of Excellence Matter and Light for Quantum Computing (ML$4$Q, Project No. EXC $2004/1$, Grant No. $390534769$), within the RTG $1995$ and within the Priority Program SPP $2244$ ``2DMP''. F.G. and R.T. are supported by the DFG through Project$-$ID $258499086$-SFB $1170$, through the W\"{u}rzburg-Dresden Cluster of Excellence on Complexity and Topology in Quantum Matter - ct.qmat Project-ID $390858490$ - EXC $2147$, and the research unit QUAST, FOR $5249-449872909$ (Project $3$). A.P. acknowledges funding from the European Research Council (ERC) under the European Union’s Horizon 2020 research and innovation program (Grant agreement No. $101002955$ - CONQUER). M.E. acknowledges funding through the DFG QUAST-FOR$5249$ - $449872909$ (Project P$6$), and through the Cluster of Excellence ”CUI: Advanced Imaging of Matter“ of the DFG – EXC $2056$ – project ID $390715994$.

\section*{Authors contributions}
F.G. conceived the project together with M.E. F.G. performed the numerical simulations. F.G. and A.P. prepared the images. F.G. wrote the first draft of the manuscript, which was then revised by M.E. with inputs from A.P., R.T. and D.M.K. All the authors contributed to the discussion of the results.

\section*{Declaration of interests}
The authors declare no competing interests.

\section*{Supplemental information}
Supplemental information can be found online at [URL].

\section*{Note added}
In Ref.~\cite{Sefidkhani2024_arXiv}, the authors also studied the effect of the fluctuations on the dynamics of the order parameter. Where overlapping, our manuscripts agree in their conclusions.


\end{document}